\documentclass[english]{article}
\usepackage[T1]{fontenc}
\usepackage[latin9]{inputenc}
\usepackage{geometry}
\usepackage{float}
\usepackage{amsmath}
\usepackage{amssymb}
\usepackage{graphicx}
\usepackage{esint}

\makeatletter

\providecommand{\tabularnewline}{\\}

\newcommand{\lyxaddress}[1]{
\par {\raggedright #1
\vspace{1.4em}
\noindent\par}
}

\usepackage{dsfont}

\makeatother

\usepackage{babel}
\begin{document}

\title{Wave propagation in relaxed micromorphic continua: modelling metamaterials
with frequency band-gaps}

\author{Angela Madeo$^{1,5}$%
\thanks{Corresponding author: angela.madeo@insa-lyon.fr%
}\quad{}, Patrizio Neff$^{2,5}$, Ionel-Dumitrel Ghiba$^{2,3}$, Luca
Placidi$^{4,5}$, Giuseppe Rosi$^{5,6}$}
\maketitle
\begin{quotation}
\begin{center}
``Sans la curiosité de l'ésprit, que serions-nous?'' Marie Curie
\par\end{center}

\bigskip{}

\end{quotation}

\lyxaddress{1. Université de Lyon-INSA, 20 Av. Albert Einstein, 69100 Villeurbanne
Cedex, France. }

\lyxaddress{2. Lehrstuhl f\"{u}r Nichtlineare Analysis und Modellierung, Fakultät
für Mathematik, Universität Duisburg-Essen, Campus Essen, Thea-Leymann
Str. 9, 45127 Essen, Germany.}

\lyxaddress{3. Department of Mathematics, University \textquotedblleft{}A.I.
Cuza\textquotedblright{}, Blvd. Carol I, no. 11, 700506 Ia\c{s}i,
Romania; Octav Mayer Institute of Mathematics, Romanian Academy, 700505
- Ia\c{s}i; Institute of Solid Mechanics, Romanian Academy, 010141-Bucharest,
Romania.}

\lyxaddress{4. Università Telematica Internazionale Uninettuno, Corso V. Emanuele
II 39, 00186 Roma, Italy.}

\lyxaddress{5. International Center M\&MOCS \textquotedblleft{}Mathematics and
Mechanics of Complex Systems\textquotedblright{}, Palazzo Caetani,
Cisterna di Latina, Italy.}

\lyxaddress{6. Laboratoire de Modelisation Multi Echelle, MSME UMR 8208 CNRS,
Université Paris-Est, 61 Avenue du Général De Gaulle, Créteil Cedex
94010, France}
\begin{abstract}
In this paper the relaxed micromorphic model proposed in \cite{NeffRelaxed ,Ghiba}
has been used to study wave propagation in unbounded continua with
microstructure. By studying dispersion relations for the considered
relaxed medium, we are able to disclose precise frequency ranges (band-gaps)
for which propagation of waves cannot occur. These dispersion relations
are strongly nonlinear so giving rise to a macroscopic dispersive
behavior of the considered medium. We prove that the presence of band-gaps
is related to a unique elastic coefficient, the so-called \textit{Cosserat
couple modulus} $\mu_{c}$, which is also responsible for the loss
of symmetry of the Cauchy force stress tensor. This parameter can
be seen as the trigger of a bifurcation phenomenon since the fact
of slightly changing its value around a given threshold drastically
changes the observed response of the material with respect to wave
propagation. We finally show that band-gaps cannot be accounted for
by classical micromorphic models as well as by Cosserat and second
gradient ones. The potential fields of application of the proposed
relaxed model are manifold, above all for what concerns the conception
of new engineering materials to be used for vibration control and
stealth technology.

\bigskip{}

\end{abstract}

\section{Introduction}

Engineering metamaterials showing exotic behaviors with respect to
wave propagation are recently attracting growing attention from the
scientific community for what concerns both modelling and experiments.
Indeed, there are multiple experimental evidences supporting the fact
that engineering microstructured materials such as phononic crystals
and lattice structures can inhibit wave propagation in particular
frequency ranges (see e.g. \cite{Vasseur,Vasseur1}). Both phononic
crystals and lattice structures are artificial materials which are
characterized by periodic microstructures in which strong contrasts
of the material properties can be observed. Also suitably ordered
granular assemblies with defects (see e.g. \cite{Merkel,Merkel1,Kafesaki})
and composites (\cite{Economou}) have shown the possibility of exhibiting
band-gaps in which wave propagation cannot occur under particular
loading conditions. In all these cases, the basic features of the
observed band-gaps are directly related to the presence of an underlying
microstructure in which strong contrasts of the elastic properties
may occur. Indeed, the rich dynamic behavior of such materials stems
mainly from their hierarchical heterogeneity at the microscopic level
which produces the mixing of the longitudinal and the transverse components
of travelling waves. It is clear that the applications of such metamaterials
would be very appealing for what concerns vibration control in engineering
structures. These materials could be used as an alternative to currently
used piezo-electric materials which are commonly employed for structural
vibration control and for this reason are widely studied in the literature
(see, among many others, \cite{Piezo1,isola2,isola6,isola1,isola3,Piezo}).

\subsection{Motivation and originality of the present work}

Different modelling efforts were recently made trying to account for
the observed band-gaps in a reliable manner. The most common models
are based on homogenization procedures on simple periodic microstructures
with strong contrast giving rise to Cauchy-type equivalent continua.
Often, it is remarked that the equivalent continua obtained by means
of these homogenization techniques are able to predict band-gaps when
the equivalent mass of the macroscopic system becomes negative (see
e.g.\cite{BoutinNegM,Huang0,Huang}). Homogenized systems with negative
equivalent mass can be successfully replaced by suitable generalized
continua. Indeed, it is known that homogenization of strongly heterogeneous
periodic systems may lead to generalized continua as equivalent macroscopic
medium (see e.g.\cite{isolaHomog}). Some rare papers (see e.g. \cite{HuangMicrostr,Huang,Vasiliev})
can be found in the literature in which lattice models with enriched
kinematics are used to model band-gaps without the need of introducing
negative equivalent masses. Nevertheless, to the authors' knowledge,
a systematic treatment of band-gap modelling based on the spirit of
micromorphic continuuum mechanics is still lacking and deserves attention.
Indeed, the idea of using generalized continuum theories to describe
microstructured materials needs to be fully developed in order to
achieve a simplified modeling and more effective conception of engineering
devices allowing to stop wave propagation in suitable frequency bands.
Continuum models are, in fact, suitable to be used as input in finite
element calculations, which are for sure one of the main basis of
engineering design.

In this paper, we propose to use the relaxed micromorphic continuum
model presented in \cite{NeffRelaxed } to study wave propagation
in microstructured materials exhibiting exotic properties with respect
to wave propagation. The used kinematics is enhanced with respect
to the one used for classical Cauchy continua by means of supplementary
kinematical fields with respect to the macroscopic displacement. The
introduced supplementary kinematical fields allow to account for the
presence of microstructure on the overall mechanical behavior of considered
continua. The considered extended kinematics has a similar form as
the one used to describe phenomena of energy trapping in internal
degrees of freedom (see e.g. \cite{Carcaterra} and references there
cited). However, the aforementioned papers deal with damped systems
differently from what done in the present paper where only conservative
systems are considered.

It is well known that the mechanical behavior of isotropic Mindlin-Eringen
media is, in general, described by means of 18 elastic constants (see
\cite{Mindlin,EringenBook,NeffRelaxed }). Nevertheless, the set of
18 parameters introduced by Mindlin in \cite{Mindlin} and Eringen
in \cite{EringenBook} is not suitable to disclose the main features
of micromorphic media mainly because of unavoidable computational
difficulties. We propose here to consider the relaxed micromorphic
model presented in \cite{NeffRelaxed } which only counts 6 parameters
and which, nevertheless, is fully able to describe the main characteristic
features of micromorphic continua. Indeed, in \cite{NeffRelaxed }
we showed that the linear isotropic microvoid model, the linear Cosserat
model and the linear microstretch model are special cases of our relaxed
micromorphic model in dislocation format. A direct identification
of the coefficients gave us that the coefficient of the Cowin-Nunziato
theory \cite{CowinNunziato}, the Mindlin-Eringen theory \cite{Mindlin,EringenBook}
and the microstretch theory \cite{EringenBook,Iesan} can be expressed
in terms of our constitutive coefficients (see also \cite{Neff constants,NeffRelaxed }).
Using these identifications one could compare all the qualitative
results obtained using our relaxed micromorphic model with those already
available in the literature (see for instance \cite{Mindlin,EringenBook,Ghiba,Ghiba1,Ghiba2,Ghiba3,Ghiba4,Ghiba5,Ghiba6}
and references therein) concerning classical theories of elastic materials
with microstructure.

By studying dispersion relations in the proposed relaxed model, we
show that wave propagation in such simplified micromorphic media is
affected by the presence of microstructure in a quite controllable
manner. More precisely, we show that, if only longitudinal waves are
allowed to propagate in the considered medium (no displacements and
micro-deformations allowed in the direction orthogonal to the direction
of wave propagation), then the presence of band-gaps can be forecasted
only with 5 parameters. The main features of these longitudinal band
gaps are seen to be directly related to the absolute rotations of
the microstructure. Nevertheless, the proposed 5-parameters relaxed
model is not able to account for the presence of band-gaps when also
transverse waves are allowed to propagate in the considered material.
In order to treat the more general case, we hence consider a slightly
generalized relaxed model by introducing only one extra elastic parameter,
the so-called \textit{Cosserat couple modulus} $\mu_{c}$, which is
related to the relative deformation of the microstructure with respect
to the macroscopic matrix. We hence end up with a relaxed micromorphic
model with only 6 parameters (5+1) which is able to account for the
prediction of band-gaps when generic waves can travel in the considered
medium.

Moreover, we consider the so-called ``classical micromorphic medium'',
i.e. a micromorphic continuum in which the whole gradient of the micro-deformation
tensor plays a role (instead of only its Curl). We show that also
the classical micromorphic continuum is not suitable to account for
the decription of band-gaps. Finally, a particular second gradient
material is obtained as limit case of the proposed classical micromorphic
continuum by letting $\mu_{e}\rightarrow\infty$ and $\mu_{c}\rightarrow\infty$.
Also in this case, we will show that the existence of band-gaps cannot
be accounted for.

We can summarize by saying that the proposed relaxed micromorphic
model (6 elastic coefficients) accounts for the description of frequency
band-gaps in microstructured media which are ``switched on'' by
the only parameter $\mu_{c}$ related to relative rotations of the
microstucture with respect to the solid macroscopic matrix. These
band gaps cannot be predicted by means of classical micromorphic or
Cosserat and second gradient media, as it will be extensively pointed
out in the body of the paper. Moreover, band gaps are impossible to
observe in the purely one-dimensional situation, since, among others,
the Curl operator vanishes in this simplified case. This absence of
band gaps is confirmed by the results found in \cite{berezowski}
in which a one-dimensional micromorphic model is presented together
with a detailed analysis of wave propagation.

We finally remark that there are some similarities between dispersion
relations for micromorphic continua (including the proposed relaxed
model) and dispersion relations of various type of plates models,
see \cite{Victor1} where Kirchhoff, Mindlin, Reissner types of plates
including consideration of rotatory inertia are investigated. In particular,
for almost all models of plates there are optical branches of dispersion
relations and linear asymptotes when the wave number tends to infinity.
On the other hand, in the theory of plates there are no band gaps
discovered within the considered models. Also porous media with strong
contrast between the solid and fluid phases show a marked dispersive
behavior (see e.g. \cite{Steeb,Steeb1}), but band gaps are not observed
in such media.

\subsection{Notations}

Let $\mathbf{x},\mathbf{y}\in\mathbb{R}^{3}$ be two vectors and $\mathbf{X},\mathbf{Y}\in\mathbb{R}^{3\times3}$
be two second order tensors of components $x_{i},y_{i}$ and $X_{ij},Y_{ij}$,
$i,j=\{1,2,3\}$, respectively. We let%
\footnote{Here and in the sequel Einstein convention of sum of repeated indices
is used if not differently specified.%
} $\text{<}\mathbf{x},\,\mathbf{y}>_{\mathbb{R}^{3}}=x_{i}y_{i}$ and
$\text{<}\mathbf{X},\,\mathbf{Y}>_{\mathbb{R}^{3\times3}}=X_{ij}Y_{ij}$
denote the scalar product on $\mathbb{R}^{3}$ and $\mathbb{R}^{3\times3}$
with associated vector norms $\left\Vert \mathbf{x}\right\Vert _{\mathbb{R}^{3}}^{2}=\text{<}\mathbf{x},\,\mathbf{x}>_{\mathbb{R}^{3}}$
and $\left\Vert \mathbf{X}\right\Vert _{\mathbb{R}^{3\times3}}^{2}=\text{<}\mathbf{X},\,\mathbf{X}>_{\mathbb{R}^{3\times3}}$,
respectively. In what follows, we omit the subscripts $\mathbb{R}^{3}$
and $\mathbb{R}^{3\times3}$ if no confusion can arise. We define
the standard divergence, gradient and curl operators for vectors and
second order tensors respectively as%
\footnote{Here and in the sequel a subscript $i$ after a comma indicates partial
derivative with respect to the space variable $X_{i}$.%
}
\begin{gather*}
\mathrm{Div}\,\mathbf{x}=x_{i,i},\quad\left(\,\nabla\,\mathbf{x}\,\right)_{ij}=x_{i,j},\quad\left(\,\mathrm{Curl}\,\mathbf{x}\,\right)_{i}=x_{a,b}\,\epsilon_{iab},\\
\\
\left(\,\mathrm{Div}\,\mathbf{X}\,\right)_{i}=X_{ij,j},\quad\left(\,\nabla\,\mathbf{X}\,\right)_{ijk}=X_{ij,k},\quad\left(\,\mathrm{Curl}\,\mathbf{X}\,\right)_{ij}=X_{ia,b}\epsilon_{jab},
\end{gather*}
where $\boldsymbol{\epsilon}$ is the standard Levi-Civita tensor.

For any second order tensor $\mathbf{X}\in\mathbb{R}^{3\times3}$
we introduce its symmetric, skew-symmetric, spheric and deviatoric
part respectively as
\[
\mathrm{sym}\:\mathbf{X}=\frac{1}{2}\left(\mathbf{X}+\mathbf{X}^{T}\right),\quad\mathrm{skew}\:\mathbf{X}=\frac{1}{2}\left(\mathbf{X}-\mathbf{X}^{T}\right),\quad\mathrm{sph}\:\mathbf{X}=\frac{1}{3}\,\mathrm{tr}\,\mathbf{X}\:\mathds1,\quad\mathrm{dev}\,\mathbf{X}=\mathbf{X}-\mathrm{sph}\:\mathbf{X},
\]
or equivalently, in index notation
\[
\negthickspace\negthickspace\negthickspace\negthickspace\negthickspace\negthickspace\negthickspace\negthickspace\negthickspace\negthickspace\left(\mathrm{sym}\:\mathbf{X}\right)_{ij}=\frac{1}{2}\left(X_{ij}+X_{ji}\right),\ \ \left(\mathrm{skew}\:\mathbf{X}\right)_{ij}=\frac{1}{2}\left(X_{ij}-X_{ji}\right),\ \ \left(\mathrm{sph}\:\mathbf{X}\right)_{ij}=\frac{1}{3}\, X_{kk}\:\delta_{ij},\ \ \left(\mathrm{dev}\,\mathbf{X}\right)_{ij}=X_{ij}-\left(\mathrm{sph}\:\mathbf{X}\right)_{ij},
\]
where $\delta_{ij}$ is the Kronecher delta tensor which is equal
to $1$ when $i=j$ and equal to $0$ if $i\neq j$. The \textit{Cartan-Lie-algebra
decomposition} for the tensor $\mathbf{X}$ is introduced as
\begin{equation}
\mathbf{X}=\mathrm{dev\: sym}\,\mathbf{X}+\mathrm{skew}\,\mathbf{X}+\mathrm{sph}\,\mathbf{X}.\label{CartanLie}
\end{equation}

\section{Equations of motion in strong form}

We describe the deformation of the considered continuum by introducing
a Lagrangian configuration $B_{L}\subset\mathbb{R}^{3}$ and a suitably
regular kinematical field $\boldsymbol{\chi}(\mathbf{X},t)$ which
associates to any material point $\mathbf{X}\in B_{L}$ its current
position $\mathbf{x}$ at time $t$. The image of the function $\boldsymbol{\chi}$
gives, at any instant $t$ the current shape of the body $B_{E}(t)$
which is often referred to as Eulerian configuration of the system.
Since we will use it in the following, we also introduce the displacement
field $\mathbf{u}(\mathbf{X},t)=\boldsymbol{\chi}(\mathbf{X},t)-\mathbf{X}$.
The kinematics of the continuum is then enriched by adding a second
order tensor field $\mathbf{P}(\mathbf{X},t)$ which accounts for
deformations associated to the microstructure of the continuum itself.
Hence, the current state of the considered continuum is identified
by 12 independent kinematical fields: 3 components of the displacement
field and 9 components of the micro-deformation field. Such a theory
of a continuum with microstructure has been derived in \cite{Mindlin}
for the linear-elastic case and re-proposed e.g. in \cite{EringenI,Eringen II,Forest0,Forest}
for the case of non-linear elasticity.

Once the used kinematics has been made clear, we can introduce the
action functional of the considered micromorphic system as
\begin{equation}
\mathcal{A}=\int_{0}^{T}\int_{B_{L}}\left(T-W\right)d\mathbf{X}\, dt
\end{equation}
where $T$ and $W$ are the kinetic and potential energies of the
considered system respectively. Denoting by $\rho$ and $\text{\ensuremath{\eta}}$
the macroscopic and microscopic mass densities respectively, we choose
the kinetic energy of the system to be
\begin{gather}
T=\frac{1}{2}\,\rho\left\Vert \,\mathbf{u}_{t}\,\right\Vert ^{2}+\frac{1}{2}\,\eta\left\Vert \,\mathbf{P}_{t}\,\right\Vert ^{2},\label{Kinetic}
\end{gather}
where we denote by a subscript $t$ partial derivative with respect
to time. We remark that, since the micro-deformation tensor $\mathbf{P}$
is dimensionless, the micro-density $\eta$ has the dimensions of
a bulk density ($Kg/m^{3}$) times the square of a length. This means
that, if $\rho'$ is the true density (per unit of macro volume) of
the material constituting the underlying microstructure of the medium
one can think to write the homogenized density $\eta$ as (see also
\cite{Mindlin})
\begin{equation}
\eta=d^{2}\:\rho',
\end{equation}
where we denoted by $d$ the characteristic length of the microscopic
inclusions%
\footnote{We remark that by considering a scalar microscopic density $\eta$,
we are limiting ourselves to cases in which the microscopic inclusions
only have one characteristic length (e.g. cubes or spheres). On the
other hand, expression (\ref{Kinetic}) for the kinetic energy can
be easily generalized by replacing the second term with $1/2\,\eta_{ij}(P_{ki})_{t}(P_{kj})_{t}$,
where the tensor $\boldsymbol{\eta}$ can be written as $\eta_{ij}=d_{ij}^{2}\rho'$.
In this way one can account for as many microscopic characteristic
lengths $d_{ij}$ as needed.%
}.

On the other hand, we will specify the choice of the strain energy
density $W$ in the next subsections by discussing the cases of the
relaxed micromorphic continuum introduced in \cite{NeffRelaxed }
and of the classical micromorphic continuum.

\subsection{The relaxed micromorphic continuum}

The strain energy density for the relaxed micromorphic continuum can
be written as

\begin{align}
W & =\mu_{e}\left\Vert \,\mathrm{sym}\left(\nabla\mathbf{u}-\mathbf{P}\right)\,\right\Vert ^{2}+\frac{\lambda_{e}}{2}\left(\mathrm{tr}\left(\nabla\mathbf{u}-\mathbf{P}\right)\right)^{2}+\mu_{h}\left\Vert \,\mathrm{sym}\,\mathbf{P}\,\right\Vert ^{2}+\frac{\lambda_{h}}{2}\left(\mathrm{tr}\,\mathbf{P}\right)^{2}\label{KinPot}\\
 & +\mu_{c}\left\Vert \,\mathrm{skew}\left(\nabla\mathbf{u}-\mathbf{P}\right)\,\right\Vert ^{2}+\frac{\alpha_{c}}{2}\left\Vert \,\mathrm{\mathrm{Cur}l}\,\mathbf{P}\,\right\Vert ^{2},\nonumber
\end{align}
where all the introduced elastic coefficients are assumed to be constant.
This decomposition of the strain energy density, valid in the isotropic,
linear-elastic case, has been proposed in \cite{NeffRelaxed ,Ghiba}
where well-posedness theorems have also been proved. It is clear that
this decomposition introduces a limited number of elastic parameters
and we will show how this may help in the physical interpretation
of these latter. Positive definiteness of the potential energy implies
the following simple relations on the introduced parameters
\begin{equation}
\mu_{e}>0,\qquad\mu_{c}>0,\qquad3\lambda_{e}+2\mu_{e}>0,\qquad\mu_{h}>0,\qquad3\lambda_{h}+2\mu_{h}>0,\qquad\alpha_{c}>0.\label{DefPos}
\end{equation}
One of the most interesting features of the proposed strain energy
density is the reduced number of elastic parameters which are needed
to fully describe the mechanical behavior of a micromorphic continuum.
Indeed, each parameter can be easily related to specific micro and
macro deformation modes. In the following, we use a strengthened set
of requirements which implies (\ref{DefPos}), namely
\begin{equation}
\mu_{e}>0,\qquad\mu_{c}>0,\qquad2\lambda_{e}+\mu_{e}>0,\qquad\mu_{h}>0,\qquad2\lambda_{h}+\mu_{h}>0,\qquad\alpha_{c}>0.\label{DefPos-1}
\end{equation}

\subsubsection{Comparison with Mindlin and Eringen models}

It can be checked that the proposed strain energy density (\ref{KinPot})
represents a particular case of the strain energy density proposed
by Mindlin (cf. Eq. (5.5) of \cite{Mindlin}). Indeed, considering
that our micro-strain tensor is the transposed of the one introduced
by Mindlin ($P_{ij}=\psi_{ji}$) and using the substitutions (cf.
also \cite{Neff_Forest_jel05}):
\begin{gather}
\mu=\mu_{h},\quad\lambda=\lambda_{h},\quad b_{1}=\lambda_{e}+\lambda_{h},\quad b_{2}=\mu_{e}+\mu_{h}+\mu_{c},\quad b_{3}=\mu_{e}+\mu_{h}-\mu_{c},\quad g_{1}=-\lambda_{h},\quad g_{2}=-\mu_{h},\label{Identification1}\\
\nonumber \\
a_{10}=\alpha_{c},\quad a_{14}=-\alpha_{c},\qquad a_{1}=a_{2}=a_{3}=a_{4}=a_{5}=a_{8}=a_{11}=a_{13}=a_{15}=0,\label{Identification2}
\end{gather}
we are able to recover that our relaxed model can be obtained as a
particular case of Mindlin's one. The relaxed strain energy density
(\ref{KinPot}) is not positive definite in the sense of Mindlin and
Eringen, but it gives rise to a well posed model (see \cite{NeffRelaxed }).
It is evident that the representation of the strain energy density
(\ref{KinPot}) is more suitable for applications than Mindlin's one
due to the reduced number of parameters. Indeed, if we consider all
the terms with the space derivatives of $\mathbf{P}$ to be vanishing
($\alpha_{c}=0$ in our model and $a_{i}=0$ in Mindlin) we have complete
equivalence of the two models when using the parameters identification
(\ref{Identification1}). We can hence start noticing that we use
in this model only 5 parameters instead of Mindlin's 7 parameters.
Things become even more interesting when looking at the terms in the
energy which involve derivatives of the microstrain tensor $\mathbf{P}$.
Indeed, our relaxed model only provides 1 extra parameter, $\alpha_{c}$,
instead of Mindlin's 11 independent parameters ($a_{1},\: a_{2},\: a_{3},\: a_{4},\: a_{5},\: a_{8},\: a_{10},\: a_{11},\: a_{13},\: a_{14},\: a_{15}$).
We claim that the proposed relaxed model is able to catch the basic
features of observable material behaviours of materials with microstructure.

When considering Eringen model for micromorphic continua (see \cite{EringenBook}
p. 273 for the strain energy density), we can recover that our relaxed
model can be obtained from Eringen's one by setting
\begin{gather}
\mu=\mu_{e}-\mu_{c},\quad\lambda=\lambda_{e},\quad\tau=\lambda_{e}+\lambda_{h},\quad\nu=-\lambda_{e},\quad\eta=\mu_{e}+\mu_{h}-\mu_{c},\quad\sigma=\mu_{c}-\mu_{e},\quad k=2\mu_{c},\label{Identification1-1}\\
\nonumber \\
\tau_{7}=\alpha_{c},\quad\tau_{9}=-\alpha_{c},\qquad\tau_{1}=\tau_{2}=\tau_{3}=\tau_{4}=\tau_{5}=\tau_{6}=\tau_{8}=\tau_{10}=\tau_{11}=0.\label{Identification2-1}
\end{gather}
Also in this case, we replace Eringen's 7 parameters with only 5 parameters
and, when considering terms with derivatives of the microstrain tensor
$\mathbf{P}$, we have only one additional parameter $\alpha_{c}$
instead of Eringen's 11 parameters.

\subsubsection{Governing equations}

Imposing the first variation of the action functional to be vanishing
(i.e. $\delta\mathcal{A}=0$), integrating by parts a suitable number
of times and considering arbitrary variations $\delta\boldsymbol{\chi}$
and $\delta\mathbf{P}$ of the basic kinematical fields, we obtain
the strong form of the bulk equations of motion of considered system
which read
\begin{align}
\rho\,\mathbf{u}_{tt} & =\mathrm{Div}\left[2\,\mu_{e}\,\mathrm{sym}\left(\nabla\mathbf{u}-\mathbf{P}\right)+\lambda_{e}\mathrm{tr}\left(\nabla\mathbf{u}-\mathbf{P}\right)\mathds1+2\,\mu_{c}\,\mathrm{skew}\left(\nabla\mathbf{u}-\mathbf{P}\right)\right],\nonumber \\
\label{eq:bulk-mod-3}\\
\eta\,\mathbf{P}_{tt} & =2\,\mu_{e}\,\mathrm{sym}\left(\nabla\mathbf{u}-\mathbf{P}\right)+\lambda_{e}\mathrm{tr}\left(\nabla\mathbf{u}-\mathbf{P}\right)\mathds1-2\,\mu_{h}\,\mathrm{sym}\,\mathbf{P}-\lambda_{h}\mathrm{tr}\,\mathbf{P}\:\mathds1\nonumber \\
 & +2\,\mu_{c}\,\mathrm{skew}\left(\nabla\mathbf{u}-\mathbf{P}\right)-\alpha_{c}\,\mathrm{\mathrm{Cur}l}\left(\mathrm{\mathrm{Cur}l}\,\mathbf{P}\right)\nonumber
\end{align}
or equivalently, in index notation%
\footnote{When using index notation for the components of the introduced tensor
fields, we will denote partial derivative with respect to time with
a superposed dot instead of a $t$ subscript.%
}
\begin{align}
\rho\,\ddot{u}_{i} & =\mu_{e}\left(u_{i,jj}-P_{ij,j}+u_{j,ij}-P_{ji,j}\right)+\lambda_{e}\left(u_{j,ji}-P_{jj,i}\right)\delta_{ij}+\mu_{c}\left(u_{i,jj}-P_{ij,j}-u_{j,ij}+P_{ji,j}\right),\nonumber \\
\label{eq:bulk-mod-3-1}\\
\eta\,\ddot{P}_{ij} & =\mu_{e}\left(u_{i,j}-P_{ij}+u_{j,i}-P_{ji}\right)+\lambda_{e}\left(u_{k,k}-P_{kk}\right)\delta_{ij}-\mu_{h}\left(P_{ij}+P_{ji}\right)-\lambda_{h}P_{kk}\delta_{ij}\nonumber \\
 & +\mu_{c}\left(u_{i,j}-P_{ij}-u_{j,i}+P_{ji}\right)+\alpha_{c}\left(P_{ij,kk}-P_{ik,jk}\right).\nonumber
\end{align}

\subsection{The classical micromorphic continuum}

From here on, we call classical micromorphic continuum a medium the
energy of which is given by
\begin{align}
W & =\mu_{e}\left\Vert \,\mathrm{sym}\left(\nabla\mathbf{u}-\mathbf{P}\right)\,\right\Vert ^{2}+\frac{\lambda_{e}}{2}\left(\mathrm{tr}\left(\nabla\mathbf{u}-\mathbf{P}\right)\right)^{2}+\mu_{h}\left\Vert \,\mathrm{sym}\,\mathbf{P}\,\right\Vert ^{2}+\frac{\lambda_{h}}{2}\left(\mathrm{tr}\,\mathbf{P}\right)^{2}\label{KinPot-1-1}\\
 & +\mu_{c}\left\Vert \,\mathrm{skew}\left(\nabla\mathbf{u}-\mathbf{P}\right)\,\right\Vert ^{2}+\frac{\alpha_{g}}{2}\left\Vert \nabla\,\mathbf{P}\right\Vert ^{2}.\nonumber
\end{align}

\subsubsection{Comparison with the Mindlin and Eringen models}

Analogously to what done for the relaxed case, we can recover that
the classical micromorphic continuum can be seen as a particular case
of Mindlin's model by means of the parameter identification (\ref{Identification1})
to which one must add:
\[
a_{10}=\alpha_{g},\qquad a_{1}=a_{2}=a_{3}=a_{4}=a_{5}=a_{6}=a_{7}=a_{8}=a_{9}=a_{11}=a_{12}=a_{13}=a_{14}=a_{15}=0.
\]
Analogously, the classical micromorphic continuum can be obtained
from Eringen's model by means of the parameter identification (\ref{Identification1-1})
to which one must add the conditions
\[
\tau_{7}=\alpha_{g},\qquad\tau_{1}=\tau_{2}=\tau_{3}=\tau_{4}=\tau_{5}=\tau_{6}=\tau_{8}=\tau_{9}=\tau_{10}=\tau_{11}=0.
\]

\subsubsection{Governing equations}

The equations of motion are the same as Eqs. (\ref{eq:bulk-mod-3}),
except for the gradient term in the second equation:
\begin{align}
\rho\,\mathbf{u}_{tt} & =\mathrm{Div}\left[2\,\mu_{e}\,\mathrm{sym}\left(\nabla\mathbf{u}-\mathbf{P}\right)+\lambda_{e}\mathrm{tr}\left(\nabla\mathbf{u}-\mathbf{P}\right)\mathds1+2\,\mu_{c}\,\mathrm{skew}\left(\nabla\mathbf{u}-\mathbf{P}\right)\right],\nonumber \\
\label{eq:bulk-mod-3-2-1}\\
\eta\,\mathbf{P}_{tt} & =2\,\mu_{e}\,\mathrm{sym}\left(\nabla\mathbf{u}-\mathbf{P}\right)+\lambda_{e}\mathrm{tr}\left(\nabla\mathbf{u}-\mathbf{P}\right)\mathds1-2\,\mu_{h}\,\mathrm{sym}\,\mathbf{P}-\lambda_{h}\mathrm{tr}\,\mathbf{P}\:\mathds1\nonumber \\
 & +2\,\mu_{c}\,\mathrm{skew}\left(\nabla\mathbf{u}-\mathbf{P}\right)+\alpha_{g}\:\mathrm{Div}\left(\nabla\mathbf{P}\right).\nonumber
\end{align}
The second equation can hence be rewritten in index notation as
\begin{align}
\eta\,\ddot{P}_{ij} & =\mu_{e}\left(u_{i,j}-P_{ij}+u_{j,i}-P_{ji}\right)+\lambda_{e}\left(u_{k,k}-P_{kk}\right)\delta_{ij}-\mu_{h}\left(P_{ij}+P_{ji}\right)-\lambda_{h}P_{kk}\delta_{ij}\label{EqBulkClass}\\
 & +\mu_{c}\left(u_{i,j}-P_{ij}-u_{j,i}+P_{ji}\right)+\alpha_{g}P_{ij,kk}.\nonumber
\end{align}

\section{Plane wave propagation in micromorphic media}

In our study of wave propagation in considered micromorphic media,
we will limit ourselves to the case of plane waves travelling in an
infinite domain. With this end in mind, we can suppose that the space
dependence of all the introduced kinematical fields is limited only
to the component $X$ of $\mathbf{X}$ which we also suppose to be
the direction of propagation of the considered wave. It is immediate
that, according to the Cartan-Lie decomposition for the tensor $\mathbf{P}$
(see Eq.  (\ref{CartanLie})), the component $P_{11}$ of the tensor
$\mathbf{P}$ itself can be rewritten as $P_{11}=P^{D}+P^{S}$ where
we set
\begin{equation}
P^{S}:=\frac{1}{3}\left(P_{11}+P_{22}+P_{33}\right),\qquad P^{D}:=\left(\mathrm{dev\: sym}\,\mathbf{P}\right)_{11}.\label{SpherDev}
\end{equation}
We also denote the components 12 and 13 of the symmetric and skew-symmetric
part of the tensor $\mathbf{P}$ respectively as
\begin{equation}
\left(\mathrm{sym}\:\mathbf{P}\right)_{1\xi}=P_{(1\xi)},\qquad\left(\mathrm{skew}\:\mathbf{P}\right)_{1\xi}=P_{\left[1\xi\right]},\quad\xi=1,2.\label{SymSkew}
\end{equation}
We finally introduce the last new variable
\begin{equation}
P^{V}=P_{22}-P_{33}.\label{VolumeVar}
\end{equation}

\subsection{The relaxed micromorphic continuum}

We want to rewrite the equations of motion (\ref{eq:bulk-mod-3-1})
in terms of the new variables (\ref{SpherDev}), (\ref{SymSkew})
and, of course, of the displacement variables $u_{i}$. Before doing
so, we introduce the quantities%
\footnote{Due to the chosen values of the parameters, which are supposed to
satisfy (\ref{DefPos-1}), all the introduced characteristic velocities
and frequencies are real. Indeed it can be checked that the condition
$\left(2\lambda_{e}+\mu_{e}\right)>0$ together with the condition
$\mu_{e}>0$, imply both the conditions $\left(3\lambda_{e}+2\mu_{e}\right)>0$
and $\left(\lambda_{e}+2\mu_{e}\right)>0$. %
}
\begin{gather}
c_{m}=\sqrt{\frac{\alpha_{c}}{\eta}},\qquad c_{s}=\sqrt{\frac{\mu_{e}+\mu_{c}}{\rho}},\qquad c_{p}=\sqrt{\frac{\lambda_{e}+2\mu_{e}}{\rho}},\nonumber \\
\nonumber \\
\omega_{s}=\sqrt{\frac{2\left(\mu_{e}+\mu_{h}\right)}{\eta}},\qquad\omega_{p}=\sqrt{\frac{\left(3\lambda_{e}+2\mu_{e}\right)+\left(3\lambda_{h}+2\mu_{h}\right)}{\eta}},\qquad\omega_{r}=\sqrt{\frac{2\mu_{c}}{\eta}},\label{Definitions}\\
\nonumber \\
\omega_{l}=\sqrt{\frac{\lambda_{h}+2\mu_{h}}{\eta}},\qquad\omega_{t}=\sqrt{\frac{\mu_{h}}{\eta}}.\nonumber
\end{gather}
With the proposed new choice of variables and considering definitions
(\ref{Definitions}) we are able to rewrite the governing equations
as different uncoupled sets of equations, namely:
\begin{itemize}
\item A set of three equations only involving longitudinal quantities:
\end{itemize}
\begin{align}
\ddot{u}_{1} & =c_{p}^{2}u_{1,11}-\frac{2\mu_{e}}{\rho}\, P_{,1}^{D}-\frac{3\lambda_{e}+2\mu_{e}}{\rho}\, P_{,1}^{S},\nonumber \\
\nonumber \\
\ddot{P}^{D} & =\frac{4}{3}\,\frac{\mu_{e}}{\eta}u_{1,1}+\frac{1}{3}c_{m}^{2}P_{,11}^{D}-\frac{2}{3}c_{m}^{2}P_{,11}^{S}-\omega_{s}^{2}P^{D},\label{Longitudinal}\\
\nonumber \\
\ddot{P}^{S} & =\frac{3\lambda_{e}+2\mu_{e}}{3\eta}u_{1,1}-\frac{1}{3}c_{m}^{2}P_{,11}^{D}+\frac{2}{3}c_{m}^{2}P_{,11}^{S}-\omega_{p}^{2}P^{S},\nonumber
\end{align}

\begin{itemize}
\item Two sets of three equations only involving transverse quantities in
the $k$-th direction, with $\xi=2,3$:
\end{itemize}
\begin{align}
\ddot{u}_{\xi} & =c_{s}^{2}u_{\xi,11}-\frac{2\mu_{e}}{\rho}\, P_{\left(1\xi\right),1}+\frac{\eta}{\rho}\omega_{r}^{2}P_{\left[1\xi\right],1},\nonumber \\
\nonumber \\
\ddot{P}_{\left(1\xi\right)} & =\frac{\mu_{e}}{\eta}u_{\xi,1}+\frac{1}{2}c_{m}^{2}P_{(1\xi)}{}_{,11}+\frac{1}{2}c_{m}^{2}P_{\left[1\xi\right],11}-\omega_{s}^{2}P_{(1\xi)},\label{Transverse}\\
\nonumber \\
\ddot{P}_{\left[1\xi\right]} & =-\frac{1}{2}\omega_{r}^{2}u_{\xi,1}+\frac{1}{2}c_{m}^{2}P_{(1\xi),11}+\frac{1}{2}c_{m}^{2}P_{\left[1\xi\right]}{}_{,11}-\omega_{r}^{2}P_{\left[1\xi\right]},\nonumber
\end{align}

\begin{itemize}
\item One equation only involving the variable $P_{\left(23\right)}$:
\end{itemize}
\begin{equation}
\ddot{P}_{\left(23\right)}=-\omega_{s}^{2}P_{\left(23\right)}+c_{m}^{2}P_{\left(23\right),11},\label{Shear}
\end{equation}

\begin{itemize}
\item One equation only involving the variable $P_{\left[23\right]}$ :
\end{itemize}
\begin{equation}
\ddot{P}_{\left[23\right]}=-\omega_{r}^{2}P_{\left[23\right]}+c_{m}^{2}P_{\left[23\right],11},\label{Rotations23}
\end{equation}

\begin{itemize}
\item One equation only involving the variable $P^{V}$:
\end{itemize}
\begin{equation}
\ddot{P}^{V}=-\omega_{s}^{2}P^{V}+c_{m}^{2}P_{,11}^{V}.\label{VolumeVariation}
\end{equation}
These 12 scalar differential equations will be used to study wave
propagation in our relaxed micromorphic media.

It can be checked that, in order to guarantee positive definiteness
of the potential energy (\ref{DefPos}) , the characteristic velocities
and frequencies introduced in Eq.(\ref{Definitions}) cannot be chosen
in a completely arbitrary way. In the numerical simulations considered
in this paper, the values of the elastic parameters are always chosen
in order to guarantee positive definiteness of the potential energy
according to Eqs.(\ref{DefPos-1}).

\subsection{The classical micromorphic continuum}

For the classical micromorphic continuum, introducing the new characteristic
velocity
\[
c_{g}=\sqrt{\frac{\alpha_{g}}{\eta}},
\]
 we get the following simplified one-dimensional equations
\begin{itemize}
\item Longitudinal
\end{itemize}
\begin{align}
\ddot{u}_{1} & =c_{p}^{2}u_{1,11}-\frac{2\mu_{e}}{\rho}\, P_{,1}^{D}-\frac{3\lambda_{e}+2\mu_{e}}{\rho}P_{,1}^{S},\nonumber \\
\nonumber \\
\ddot{P}^{D} & =\frac{4}{3}\,\frac{\mu_{e}}{\eta}u_{1,1}+c_{g}^{2}P_{,11}^{D}-\omega_{s}^{2}P^{D},\label{Longitudinal-1}\\
\nonumber \\
\ddot{P}^{S} & =\frac{3\lambda_{e}+2\mu_{e}}{3\eta}u_{1,1}+c_{g}^{2}P_{,11}^{S}-\omega_{p}^{2}P^{S},\nonumber
\end{align}

\begin{itemize}
\item Two sets of three equations only involving transverse quantities in
the $k$-th direction, with $\xi=2,3$:
\end{itemize}
\begin{align}
\ddot{u}_{\xi} & =c_{s}^{2}u_{\xi,11}-\frac{2\mu_{e}}{\rho}\, P_{\left(1\xi\right),1}+\frac{\eta}{\rho}\omega_{r}^{2}P_{\left[1\xi\right],1},\nonumber \\
\nonumber \\
\ddot{P}_{\left(1\xi\right)} & =\frac{\mu_{e}}{\eta}\, u_{\xi,1}+c_{g}^{2}P_{(1\xi)}{}_{,11}-\omega_{s}^{2}P_{(1\xi)},\label{Transverse-1}\\
\nonumber \\
\ddot{P}_{\left[1\xi\right]} & =-\frac{1}{2}\omega_{r}^{2}u_{\xi,1}+c_{g}^{2}P_{\left[1\xi\right]}{}_{,11}-\omega_{r}^{2}P_{\left[1\xi\right]},\nonumber
\end{align}

\begin{itemize}
\item One equation only involving the variable $P_{\left(23\right)}$:
\end{itemize}
\begin{equation}
\ddot{P}_{\left(23\right)}=-\omega_{s}^{2}P_{\left(23\right)}+c_{g}^{2}P_{\left(23\right),11},\label{Shear-1}
\end{equation}

\begin{itemize}
\item One equation only involving the variable $P_{\left[23\right]}$ :
\end{itemize}
\begin{equation}
\ddot{P}_{\left[23\right]}=-\omega_{r}^{2}P_{\left[23\right]}+c_{g}^{2}P_{\left[23\right],11},\label{Rotations23-1}
\end{equation}

\begin{itemize}
\item One equation only involving the variable $P^{V}$:
\end{itemize}
\begin{equation}
\ddot{P}^{V}=-\omega_{s}^{2}P^{V}+c_{g}^{2}P_{,11}^{V}.\label{VolumeVariation-1}
\end{equation}

\section{Linear waves in micromorphic media}

In this section we study the dispersion relations for the considered
relaxed micromorphic continuum, as well as for the Classical micromorphic
continuum. We also consider dispersion relations for Cosserat media
obtained as a degenerate limit case of our relaxed model. Finally,
dispersion relations for a second gradient continuum obtained as a
limit case of the classic micromorphic continuum are also presented.
It will be pointed out that only our relaxed micromorphic model can
disclose the presence of band-gaps.

\subsection{Micro-oscillations}

We start studying a particular solution of the set of introduced differential
equations by setting
\begin{gather*}
u_{i}=0,\ i=1,2,3,\quad P_{(23)}=Re\left\{ \alpha_{(23)}e^{i\omega t}\right\} ,\quad P_{[23]}=Re\left\{ \alpha_{[23]}e^{i\omega t}\right\} ,\quad P^{V}=Re\left\{ \alpha^{V}e^{i\omega t}\right\} ,\\
\\
P_{(1\xi)}=Re\left\{ \alpha_{(1\xi)}e^{i\omega t}\right\} ,\quad P_{[1\xi]}=Re\left\{ \alpha_{[1\xi]}e^{i\omega t}\right\} ,\ \xi=2,3.
\end{gather*}
Replacing these expressions for the unknown variables in each of Eqs.(\ref{Longitudinal})-(\ref{VolumeVariation}),
noticing that the space derivatives are vanishing, one gets that the
first of Eqs.(\ref{Longitudinal}) and (\ref{Transverse}) are such
that the frequency calculated for the waves $u_{1}$ and $u_{\xi}$
is vanishing, i.e.
\begin{equation}
\omega=0.\label{Acoustic}
\end{equation}
Moreover, from the remaining equations one gets the values of frequency
$\text{\ensuremath{\omega}}$ for the waves $P^{D}$, $P^{S}$, $P_{(1\xi)}$,
$P_{[1\xi]}$, $P_{(23)}$, $P_{[23]}$ and $P^{V}$ respectively
\begin{equation}
\omega^{2}=\omega_{s}^{2},\quad\omega^{2}=\omega_{p}^{2}\quad\omega^{2}=\omega_{s}^{2},\quad\omega^{2}=\omega_{r}^{2},\quad\omega^{2}=\omega_{s}^{2},\quad\omega^{2}=\omega_{r}^{2},\quad\omega^{2}=\omega_{s}^{2}.\label{Optic}
\end{equation}
It is clear that one gets exactly the same result when considering
the complete Mindlin-Eringen model, since the only difference is on
the space derivatives of $\mathbf{P}$ which do not intervene when
studying micro-oscillations. The characteristic values of the frequencies
given in Eq.(\ref{Optic}) are fixed once the material parameters
of the system are specified (see Eqs.(\ref{Definitions})) and they
represent the limit values for $k\rightarrow0$ of the eigenvalues
$\omega(k)$ associated to propagative waves, as it will be better
shown in the next section. This preliminary study of micro-oscillations
allows a precise classification of waves which can propagate in a
micromorphic medium. More particularly, we can distinguish two types
of propagative waves:
\begin{itemize}
\item \textbf{acoustic waves}, i.e. waves which have vanishing frequency
when the wavenumber $k$ is vanishing,
\item \textbf{optic waves,} i.e. waves which have non-vanishing frequency
when the wavenumber $k$ is vanishing.
\end{itemize}
Indeed, we will see in the following, that a third type of wave may
exist in relaxed micromorphic media under suitable hypothesis on the
constitutive parameters:
\begin{itemize}
\item \textbf{standing (or evanescent) waves,} i.e. waves which have imaginary
wavenumber $k$ corresponding to some frequency ranges. These waves
do not propagate, but keep oscillating in a given, limited region
of space.
\end{itemize}
According to the performed study of micro-oscillations we can conclude
that
\begin{itemize}
\item For the \textbf{uncoupled waves} $P_{(23)}$, $P_{[23]}$ and $P^{V}$,
only optic waves are found with cutoff frequencies $\omega_{s}$,
$\omega_{r}$ and $\omega_{s}$. Moreover, the structure of the governing
equations for $P_{(23)}$ and $P^{V}$ are formally identical (see
Eqs. (\ref{Shear}) and (\ref{VolumeVariation})) so that the associated
dispersion relations will give rise to superimposed curves. In the
limit $\mu_{c}\rightarrow0$ one has from Eq. (\ref{Definitions})
that $\omega_{r}\rightarrow0.$ This means that in this limit case
one has one acoustic waves and two superimposed optic waves with cutoff
frequency $\omega_{s}$.
\item For the\textbf{ longitudinal waves} $u_{1}$, $P^{D}$, $P^{S}$,
we can identify one acoustic wave and two optic waves with cutoff
frequencies $\omega_{s}$ and $\omega_{p}$. In the limit $\mu_{c}\rightarrow0$,
which implies that $\omega_{r}\rightarrow0$, the situation for longitudinal
waves remains unchanged.
\item For the \textbf{transverse waves} $u_{\xi}$, $P_{(1\xi)}$, $P_{[1\xi]}$,
($\xi=1,2$) we identify one acoustic wave and two optic waves with
cutoff frequencies $\omega_{s}$ and $\omega_{r}$. In the limit $\mu_{c}\rightarrow0$
which implies that $\omega_{r}\rightarrow0$, we have two acoustic
waves and one optic wave with cutoff frequency $\omega_{s}$.
\end{itemize}

\subsection{Planar wave propagation in the relaxed micromorphic continuum}

We now look for a wave form solution of the previously derived equations
of motion. We start from the uncoupled equations (\ref{Shear})-(\ref{VolumeVariation})
and assume that the involved unknown variables take the harmonic form
\begin{equation}
P_{\left(23\right)}=Re\left\{ \beta_{\left(23\right)}e^{i(kX-\omega t)}\right\} ,\quad P_{\left[23\right]}=Re\left\{ \beta_{\left[23\right]}e^{i(kX-\omega t)}\right\} ,\quad P^{V}=Re\left\{ \beta^{V}e^{i(kX-\omega t)}\right\} ,\label{WaveForm1}
\end{equation}
where $\beta_{\left(23\right)}$, $\beta_{\left[23\right]}$ and $\beta^{V}$
are the amplitudes of the three introduced waves. It can be remarked
that the variables $P_{\left(23\right)}$, $P_{\left[23\right]}$
and $P^{V}$ respectively represent transverse (with respect to wave
propagation) micro-shear, transverse micro-rotation and transverse
micro-deformations at constant volume. Replacing this wave form in
Eqs. (\ref{Shear})-(\ref{VolumeVariation}) and simplifying one obtains
the following dispersion relations respectively:
\begin{equation}
\omega(k)=\sqrt{\omega_{s}^{2}+k^{2}c_{m}^{2}},\qquad\omega(k)=\sqrt{\omega_{r}^{2}+k^{2}c_{m}^{2}},\qquad\omega(k)=\sqrt{\omega_{s}^{2}+k^{2}c_{m}^{2}}.\label{Eigen1}
\end{equation}
We notice that for a vanishing wave number ($k=0$) the dispersion
relations for the three considered waves give non-vanishing frequencies
which correspond to the ones calculated in the previous subsection
for the same waves. This is equivalent to say that the waves associated
to the three considered variables $P_{\left(23\right)}$, $P_{\left[23\right]}$
and $P^{V}$ are so-called optic waves. Moreover, we also notice that
the dispersion relation for the variables $P_{\left(23\right)}$ and
$P^{V}$ are the same: this means that the wave-form solutions for
these variables coincide modulo a scalar multiplication factor (see
Eqs. (\ref{WaveForm1})).

We now want to study harmonic solutions for the differential systems
(\ref{Longitudinal}) and (\ref{Transverse}). To do so, we introduce
the unknown vectors $\mathbf{v}_{1}=\left(u_{1},P^{D},P^{S}\right)$
and $\mathbf{v}_{\xi}=\left(u_{\xi},P_{(1\xi)},P_{[1\xi]}\right),\ \xi=2,3$
and look for wave form solutions of equations (\ref{Longitudinal})
and (\ref{Transverse}) in the form
\begin{equation}
\mathbf{v}_{1}=Re\left\{ \boldsymbol{\beta}e^{i(kX-\omega t)}\right\} ,\qquad\mathbf{v}_{\xi}=Re\left\{ \boldsymbol{\gamma}^{\xi}e^{i(kX-\omega t)}\right\} ,\ \xi=2,3\label{WaveForm2}
\end{equation}
where $\boldsymbol{\beta}=(\beta_{1},\beta_{2},\beta_{3})^{T}$ and
$\boldsymbol{\gamma}^{\xi}=(\gamma_{1}^{\xi},\gamma_{2}^{\xi},\gamma_{3}^{\xi})^{T}$
are the unknown amplitudes of considered waves. Replacing this expressions
in equations (\ref{Longitudinal}) and (\ref{Transverse}) one gets
respectively
\begin{equation}
\mathbf{A}_{1}\cdot\boldsymbol{\beta}=0,\qquad\mathbf{A}_{\xi}\cdot\boldsymbol{\gamma}^{\xi}=0,\qquad\xi=2,3,\label{AlgSys}
\end{equation}
where
\begin{gather*}
\mathbf{A}_{1}=\left(\begin{array}{ccc}
-\omega^{2}+c_{p}^{2}\, k^{2} & \, i\: k\:2\mu_{e}/\rho\  & i\: k\:\left(3\lambda_{e}+2\mu_{e}\right)/\rho\\
\\
-i\: k\,\frac{4}{3}\,\mu_{e}/\eta & -\omega^{2}+\frac{1}{3}k^{2}c_{m}^{2}+\omega_{s}^{2} & -\frac{2}{3}\, k^{2}c_{m}^{2}\\
\\
-\frac{1}{3}\, i\, k\:\left(3\lambda_{e}+2\mu_{e}\right)/\eta & -\frac{1}{3}\, k^{2}\, c_{m}^{2} & -\omega^{2}+\frac{2}{3}\, k^{2}\, c_{m}^{2}+\omega_{p}^{2}
\end{array}\right),\\
\\
\\
\mathbf{A}_{2}=\mathbf{A}_{3}=\left(\begin{array}{ccc}
-\omega^{2}+k^{2}c_{s}^{2}\  & \, i\, k\,2\mu_{e}/\rho\  & -i\, k\,\frac{\eta}{\rho}\omega_{r}^{2},\\
\\
-\, i\, k\,2\mu_{e}/\eta, & -2\omega^{2}+k^{2}c_{m}^{2}+2\omega_{s}^{2} & k^{2}c_{m}^{2}\\
\\
i\, k\,\omega_{r}^{2} & k^{2}c_{m}^{2} & -2\omega^{2}+k^{2}c_{m}^{2}+2\omega_{r}^{2}
\end{array}\right).
\end{gather*}
In order to have non-trivial solutions of the algebraic systems (\ref{AlgSys}),
one must impose that
\begin{equation}
\mathrm{det}\,\mathbf{A}_{1}=0,\qquad\mathrm{det}\,\mathbf{A}_{2}=0,\qquad\mathrm{det}\,\mathbf{A}_{3}=0,\label{Dispersion}
\end{equation}
which allow us to determine so-called dispersion relations $\omega=\omega\left(k\right)$
for the longitudinal and transverse waves in the relaxed micromorphic
conyinuum. As it will be better explained in the next section, the
eigenvalues $\omega(k)$ solutions of (\ref{Dispersion}) are associated
both to optic waves and to acoustic waves.

\subsection{Planar wave propagation in the classical micromorphic continuum}

We want to deduce in this subsection the dispersion relations for
the classical micromorphic model, i.e. for the dispersion relations
associated to the energy (\ref{KinPot-1-1}). If we replace the wave-form
solution (\ref{WaveForm2}) in the longitudinal and transverse equations
(\ref{Longitudinal-1}) and (\ref{Transverse-1}) we get
\begin{equation}
\mathbf{B}_{1}\cdot\boldsymbol{\beta}=0,\qquad\mathbf{B}_{\xi}\cdot\boldsymbol{\gamma}^{\xi}=0,\qquad\xi=2,3,\label{AlgSys-1}
\end{equation}
where
\begin{gather*}
\mathbf{B}_{1}=\left(\begin{array}{ccc}
-\omega^{2}+c_{p}^{2}\, k^{2} & \, i\: k\:2\mu_{e}/\rho\  & i\: k\:\left(3\lambda_{e}+2\mu_{e}\right)/\rho\\
\\
-i\: k\,\frac{4}{3}\mu_{e}/\eta & -\omega^{2}+k^{2}c_{g}^{2}+\omega_{s}^{2} & 0\\
\\
-\frac{1}{3}\, i\, k\:\left(3\lambda_{e}+2\mu_{e}\right)/\eta & 0 & -\omega^{2}+k^{2}\, c_{g}^{2}+\omega_{p}^{2}
\end{array}\right),\\
\\
\\
\mathbf{B}_{2}=\mathbf{B}_{3}=\left(\begin{array}{ccc}
-\omega^{2}+k^{2}c_{s}^{2}\  & \, i\, k\,2\mu_{e}/\rho\  & -i\, k\,\frac{\eta}{\rho}\omega_{r}^{2},\\
\\
-\, i\, k\,2\mu_{e}/\eta, & -2\omega^{2}+2\, k^{2}c_{g}^{2}+2\omega_{s}^{2} & 0\\
\\
i\, k\,\omega_{r}^{2} & 0 & -2\omega^{2}+2\, k^{2}c_{g}^{2}+2\omega_{r}^{2}
\end{array}\right).
\end{gather*}
In order to have non-trivial solutions of the algebraic systems (\ref{AlgSys-1}),
one must impose that
\begin{equation}
\mathrm{det}\,\mathbf{B}_{1}=0,\qquad\mathrm{det}\,\mathbf{B}_{2}=0,\qquad\mathrm{det}\,\mathbf{B}_{3}=0,\label{Dispersion-1}
\end{equation}
which allow us to determine so-called dispersion relations $\omega=\omega\left(k\right)$
for the longitudinal and transverse waves in the classical micromorphic
continuum.

As for the uncoupled waves, the dispersion relations associated to
Eqs.(\ref{Shear-1}),(\ref{Rotations23-1}) and (\ref{VolumeVariation-1})
are respectively:
\begin{equation}
\omega(k)=\sqrt{\omega_{s}^{2}+k^{2}c_{g}^{2}},\qquad\omega(k)=\sqrt{\omega_{r}^{2}+k^{2}c_{g}^{2}},\qquad\omega(k)=\sqrt{\omega_{s}^{2}+k^{2}c_{g}^{2}}.\label{Eigen1-1}
\end{equation}

\section{\label{NumericalSim}The relaxed micromorphic model: numerical results }

In this section, following Mindlin \cite{Mindlin,EringenBook}, we
will show the dispersion relations $\omega=\omega(k)$ associated
to the considered relaxed micromorphic model. The analysis of dispersion
relations in \cite{Mindlin,EringenBook} was of qualitative nature,
due to the huge number of constitutive parameters (18 elastic coefficients)
which are assumed to be non-vanishing and to computational difficulties.
On the other hand, thanks to the relaxed constitutive assumption (\ref{KinPot}),
we are able to clearly show precise dispersion relations for the considered
cases and to associate to few constitutive parameters the main mechanisms
associated to wave propagation in micromorphic media. As principal
obtained result, we will clearly point out that band-gaps can be forecast
only when considering a non-vanishing Cosserat couple modulus $\mu_{c}$
in our relaxed micromorphic model. This parameter is associated to
the relative rotation of the microstructure with respect to the continuum
matrix.
\begin{table}[H]
\begin{centering}
\begin{tabular}{|c|c|c|}
\hline
Parameter & Value & Unit\tabularnewline
\hline
\hline
$\mu_{e}$ & $200$ & $MPa$\tabularnewline
\hline
$\lambda_{e}=2\mu_{e}$ & 400 & $MPa$\tabularnewline
\hline
$\mu_{c}=2.2\mu_{e}$ & $ $440 & $MPa$\tabularnewline
\hline
$\mu_{h}$ & 100 & $MPa$\tabularnewline
\hline
$\lambda_{h}$ & $100$ & $MPa$\tabularnewline
\hline
$L_{c}\ $  & $3$ & $mm$\tabularnewline
\hline
$L_{g}\ $  & $3$ & $mm$\tabularnewline
\hline
$\alpha_{c}=\mu_{e}L_{c}^{2}$ & $1.8\times10^{-3}$ & $MPa\: m^{2}$\tabularnewline
\hline
$\alpha_{g}=\mu_{e}L_{g}^{2}$ & $1.8\times10^{-3}$ & $MPa\: m^{2}$\tabularnewline
\hline
$\rho$ & $2000$ & $Kg/m^{3}$\tabularnewline
\hline
$\rho$' & $2500$ & $Kg/m^{3}$\tabularnewline
\hline
$d$ & $2$ & $mm$\tabularnewline
\hline
$\eta=d^{2}\,\rho'$ & $ $$10^{-2}$ & $Kg/m$\tabularnewline
\hline
\end{tabular}\quad{}\quad{}\quad{}\quad{}%
\begin{tabular}{|c|c|c|}
\hline
Parameter & Value & Unit\tabularnewline
\hline
\hline
$\lambda$ & $82.5$ & $MPa$\tabularnewline
\hline
$\mu$ & $66.7$ & $MPa$\tabularnewline
\hline
$E$ & $170$ & $MPa$\tabularnewline
\hline
$\nu$ & $0.28$ & $-$\tabularnewline
\hline
\end{tabular}
\par\end{centering}

\caption{\label{ParametersValues}Values of the parameters of the relaxed model
used in the numerical simulations (left) and corresponding values
of the Lamé parameters and of the Young modulus and Poisson ratio
(right).}

\end{table}
 We start by showing in Tab.\ref{ParametersValues} (left) the values
of the parameters of the relaxed model used in the performed numerical
simulations. In order to make the obtained results more exploitable,
we also recall that in \cite{NeffRelaxed } the following homogenized
formulas were obtained which relate the parameters of the relaxed
model to the macroscopic Lamé parameters $\lambda$ and $\mu$ which
are usually measured by means of standard mechanical tests
\begin{equation}
\mu_{e}=\frac{\mu_{h}\,\mu}{\mu_{h}-\mu},\qquad2\mu_{e}+3\lambda_{e}=\frac{(2\mu_{h}+3\lambda_{h})\left(2\mu+3\lambda\right)}{(2\mu_{h}+3\lambda_{h})-\left(2\mu+3\lambda\right)}.\label{Homogenized}
\end{equation}
These relationships imply that the following inequalities are satisfied
\[
\mu_{h}>\mu,\qquad3\lambda_{h}+2\mu_{h}>3\lambda+2\mu.
\]
It is clear that, once the values of the parameters of the relaxed
models are known, the standard Lamé parameters can be calculated by
means of formulas (\ref{Homogenized}), which is what was done in
Tab.\ref{ParametersValues} (right). To the sake of completeness,
we also show in the same table the corresponding Young modulus and
Poisson ratio, calculated by means of the standard formulas
\begin{equation}
E=\frac{\mu\left(3\lambda+2\mu\right)}{\lambda+\mu},\qquad\text{\ensuremath{\nu}}=\frac{\lambda}{2\left(\lambda+\mu\right)}.
\end{equation}

\subsection{The relaxed micromorphic model with $\mu_{c}=0$.}

\begin{figure}[H]
\centering{}%
\begin{tabular}{ccccc}
\includegraphics[scale=0.7]{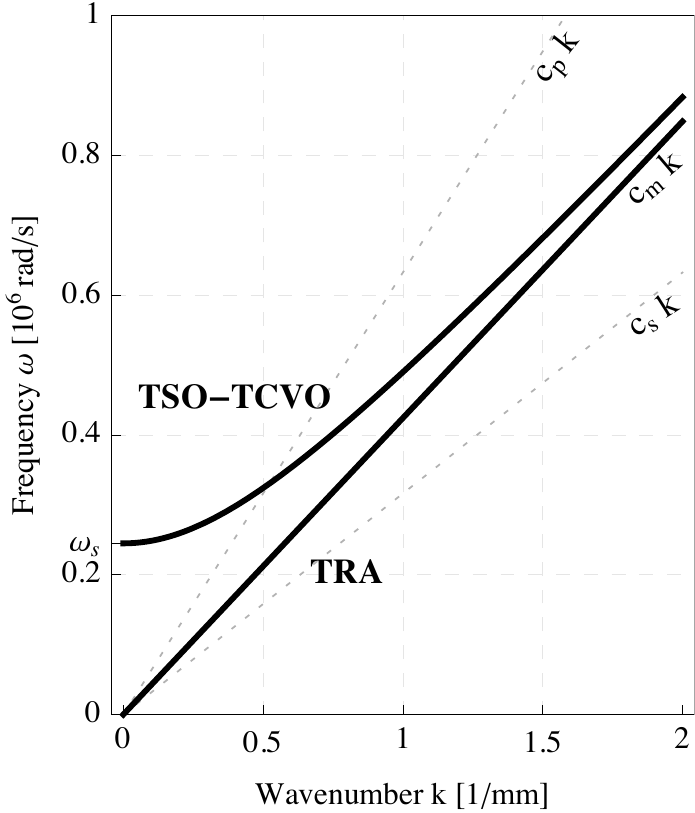}  & \includegraphics[scale=0.7]{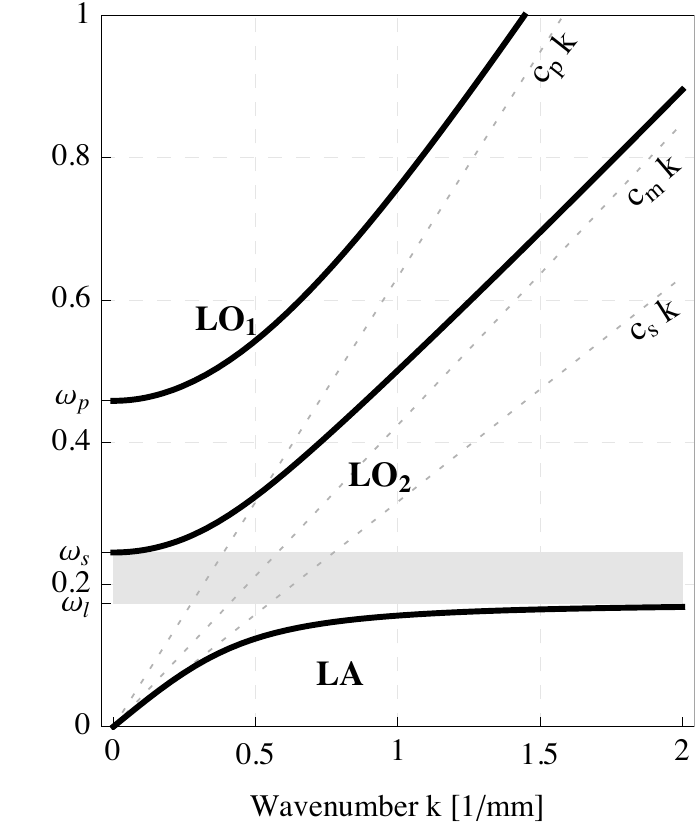} & \includegraphics[scale=0.7]{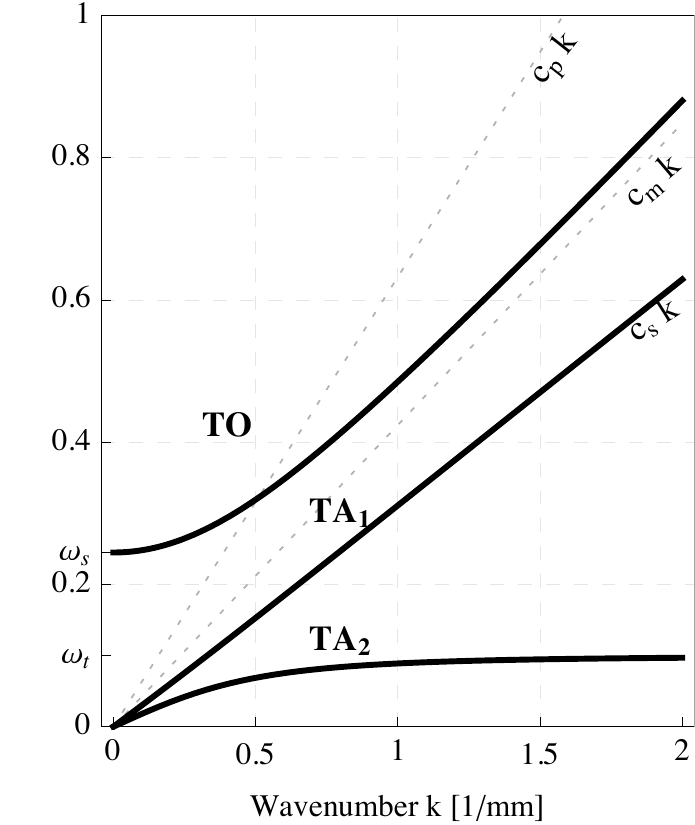} &  & \tabularnewline
(a) & (b) & (c) &  & \tabularnewline
\end{tabular}\caption{\label{DispersionDiag}Dispersion relations $\omega=\omega(k)$ for
the relaxed model with vanishing Cosserat couple modulus ($\mu_{c}=0$):
uncoupled waves (a), longitudinal waves (b) and transverse waves (c).
TRA: transverse rotational acoustic, TSO: transverse shear optic,
TCVO: transverse constant-volume optic, LA: longitudinal acoustic,
LO$_{1}$-LO$_{2}$: first and second longitudinal optic, TO: transverse
optic, TA$_{1}$-TA$_{2}$: first and second transverse acoustic.}
\end{figure}

We start by showing the dispersion relations of the algebraic problem
(\ref{Dispersion}) for the particular case $\mu_{c}=0$. Figure \ref{DispersionDiag}
shows separately the behaviors of the uncoupled waves $P_{\left(23\right)}$,
$P_{\left[23\right]}$ and $P^{V}$ (Fig. \ref{DispersionDiag}(a)),
of the longitudinal waves $u_{1},\, P^{D},\, P^{S}$ (Fig. \ref{DispersionDiag}(b))
and of the transverse waves $u_{\xi},P_{(1\xi)},P_{[1\xi]}$ (Fig.
\ref{DispersionDiag}(c)).

We can recover from Fig. \ref{DispersionDiag}(a) one acoustic, rotational
wave ($TRA$) and two superimposed optic waves ($TSO$ and $TCVO$)
with cutoff frequency $\omega_{s}$. The acoustic wave shows a non-dispersive
behavior (the dispersion curve is a straight line). This implies that,
when considering relaxed micromorphic media with $\mu_{c}=0$, there
is always at least one wave for which the wavenumber is always real
independently of the value of frequency. This fact guarantees wave
propagation inside the considered medium for all frequency ranges
(no global band-gaps).

Figure \ref{DispersionDiag}(b) shows that longitudinal waves indeed
involve one acoustic wave ($LA$) and two optic waves ($LO_{1}$ and
$LO_{2}$) with cutoff frequencies $\omega_{p}$ and $\omega_{s}$
respectively. Moreover, it is found that the acoustic wave has an
horizontal asymptote at $\omega=\omega_{l}$. As a consequence, it
can be seen that a frequency range $\left(\omega_{s},\omega_{l}\right)$
exists in which the wavenumber becomes imaginary for longitudinal
waves. According to Eqs. (\ref{WaveForm2}), an imaginary wavenumber
$k$ gives rise to solutions which are exponentials decaying in space.
Waves of this type are so-called standing waves which do not propagate,
but keep oscillating in a limited region of space.

Finally, figure \ref{DispersionDiag}(c) confirms that, for transverse
waves, two acoustic waves ($TA{}_{1}$ and $TA{}_{2}$) and one optic
wave ($TO$) with cutoff frequency $\omega_{s}$ can be identified.
One of the acoustic waves has an horizontal asymptote at $\omega=\omega_{t}$.
It is easy to recognize that, due to the existence of the non-dispersive,
transverse, acoustic wave $TA_{1}$, there always exists at least
one propagative wave for any chosen value of the frequency.

We can conclude that, in general, when considering the relaxed micromorphic
medium as a whole (all the 12 waves), there always exist waves which
propagate inside the considered medium independently of the value
of frequency. On the other hand, if one considers a particular case
(obtained by imposing suitable kinematical constraints) in which only
longitudinal waves can propagate, then in the frequency range $\left(\omega_{s},\omega_{l}\right)$
only standing wave exist which do not allow for wave propagation.
In this very particular case, the frequency range $\left(\omega_{s},\omega_{l}\right)$
can be considered as a band-gap for longitudinal waves. According
to definitions given in Eq. (\ref{Definitions}) the depth of this
frequency band is controlled by the three parameters $\mu_{e}$, $\mu_{h}$
and $\lambda_{h}$.

\subsection{The relaxed micromorphic model with $\mu_{c}>0$.}

\begin{figure}[H]
\begin{centering}
\begin{tabular}{ccccc}
\includegraphics[scale=0.7]{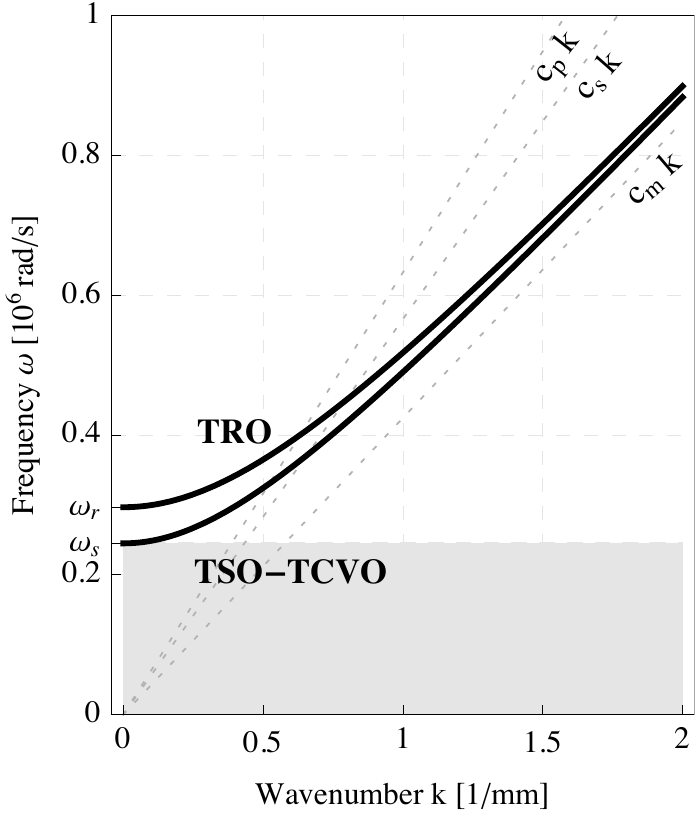}  & \includegraphics[scale=0.7]{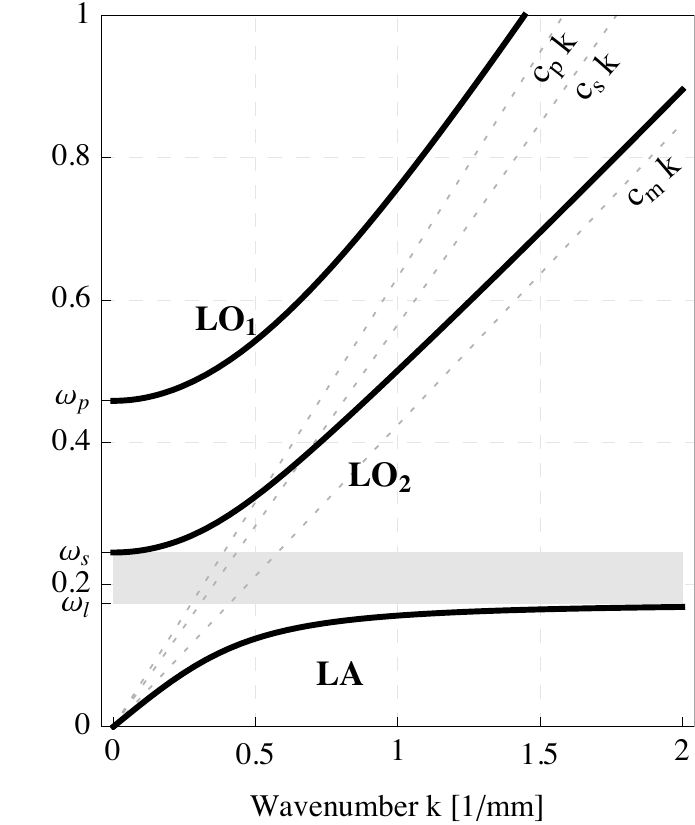} & \includegraphics[scale=0.7]{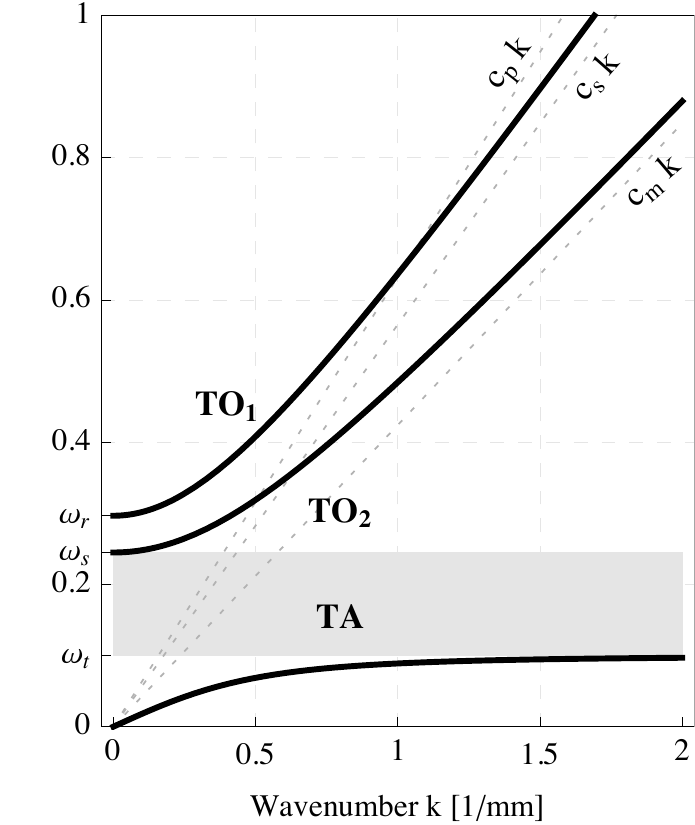} &  & \tabularnewline
(a) & (b) & (c) &  & \tabularnewline
\end{tabular}
\par\end{centering}

\caption{\label{separate}Dispersion relations $\omega=\omega(k)$ for the
relaxed model with non-vanishing Cosserat couple modulus ($\mu_{c}>0$):
uncoupled waves (a), longitudinal waves (b) and transverse waves (c).
TRO: transverse rotational optic, TSO: transverse shear optic, TCVO:
transverse constant-volume optic, LA: longitudinal acoustic, LO$_{1}$-LO$_{2}$:
first and second longitudinal optic, TA: transverse acoustic, TO$_{1}$-TO$_{2}$:
first and second transverse optic.}

\end{figure}

In this subsection we discuss the behavior with respect to wave propagation
of relaxed micromorphic continua in the general case in which the
Cosserat couple modulus $\mu_{c}$ is assumed to be non-vanishing.
We start by showing the dispersion relations in figure \ref{separate}.
As for the uncoupled waves we have, in this case, only optic waves:
two superimposed ($TSO$ and $TCVO$) with cutoff frequency $\omega_{s}$
and one ($TRO$) with cutoff frequency $\omega_{r}$ (see Fig. \ref{separate}(a)).
When considering longitudinal waves (see Fig. \ref{separate}(b))
the situation is unchanged with respect to the previous case ($\mu_{c}=0$)
and one can observe one acoustic wave ($LA$) and two optic waves
($LO{}_{1}$ and $LO_{2}$) with cutoff frequencies $\omega_{p}$
and $\omega_{s}$ respectively. If we finally consider the transverse
waves (Fig. \ref{separate}(c)), we can remark that there exists one
acoustic wave ($TA$ ) and two optic waves ($TO{}_{1}$ and $TO_{2}$)
with cutoff frequencies $\omega_{s}$ and $\omega_{r}$ respectively.
It can be easily noticed that, in all three cases, there exist frequency
ranges in which no propagative wave can be found. This means that
the wavenumber becomes imaginary and only standing waves exist. For
the situation depicted in Fig. \ref{separate}, the frequency ranges
for which only standing waves exist are $\left(0,\omega_{s}\right)$
for uncoupled waves, $\left(\omega_{l},\omega_{s}\right)$ for longitudinal
waves and $\left(\omega_{t},\omega_{s}\right)$ for transverse waves.
The intersection of these three intervals being non-empty, we can
conclude that a frequency band-gap $\left(\omega_{l},\omega_{s}\right)$
exists in the considered relaxed micromorphic medium corresponding
to which any type of wave can propagate independently of the value
of $k$. More precisely, if one imagines to excite the considered
medium with a signal the frequency of which falls in the range $(\omega_{l},\omega_{s})$,
this signal cannot propagate inside the considered medium but it keep
oscillating close to the point of application of the initial condition.
It is clear that this feature is a fundamental tool to conceive micro-structured
materials which can be used as pass- and stop- bands. Investigations
in this sense would bring new insights towards high-tech solutions
in the field of vibration control and will be the object of forthcoming
studies. It is evident (see Eqs. (\ref{Definitions})) that, in general,
the relative positions of the horizontal asymptotes $\omega_{l}$
and $\omega_{t}$ as well as of the cutoff frequencies $\omega_{s}$,
$\omega_{r}$ and $\omega_{p}$ can vary depending on the values of
the constitutive parameters. In particular, we can observe that the
relative position of the characteristic frequencies defined in (\ref{Definitions})
can vary depending on the values of the constitutive parameters. It
can be checked that, in the case in which $\lambda_{e}>0$ and $\lambda_{h}>0$
one always has $\omega_{p}>\omega_{s}>\omega_{t}$ and $\omega_{l}>\omega_{t}$.
The relative position of $\omega_{l}$ and of $\omega_{s}$ can vary
depending on the values of the parameters $\lambda_{h}$ and $\mu_{h}$.

It is easy to verify that one can have band-gaps for longitudinal
waves only if the horizontal asymptote $\omega_{l}$ is such that
$\omega_{s}>\omega_{l}$ (which implies $\lambda_{h}<2\mu_{e}$).
As for transverse waves, it can be checked that band-gaps can exist
only if $\omega_{r}>\omega_{t}$ (which implies $\mu_{c}>\mu_{h}/2$).
A band gap for the uncoupled waves always exist (independently of
the values of the constitutive parameters) for frequencies between
$0$ and the smaller among $\omega_{r}$ and $\omega_{s}$. It is
clear that some stronger conditions are needed in order to have a
global band gap which do not allow for any kind of waves (transverse,
longitudinal and uncoupled) to propagate inside the considered microstructured
material. More particularly, it can be checked that, in order to have
a global band gap, the following conditions must be simultaneously
satisfied
\[
\omega_{s}>\omega_{l}\quad\textrm{and}\quad\omega_{r}>\omega_{l}.
\]
In terms of the constitutive parameters of the relaxed model, we can
say that global band-gaps can exist, in the case in which one considers
positive values for the parameters $\lambda_{e}$ and $\lambda_{h}$
, if and only if we have simultaneously
\begin{gather}
\nonumber \\
0<\mu_{e}<+\infty,\qquad0<\lambda_{h}<2\mu_{e},\qquad\mu_{c}>\frac{\lambda_{h}+2\text{\ensuremath{\mu}}_{h}}{2}.\label{Band Gaps}\\
\nonumber
\end{gather}
As far as negative values for $\lambda_{e}$ and $\lambda_{h}$ are
allowed, the conditions for band gaps are not so straightforward as
(\ref{Band Gaps}). We do not consider this possibility in this paper,
leaving this point open for further considerations.

We can conclude by saying that, the fact of switching on a unique
parameter, namely the Cosserat couple modulus $\mu_{c}$, allows for
the description of frequency band-gaps in which no propagation can
occur. This parameter can hence be seen as a discreteness quantifier
which starts accounting for lattice discreteness as far as it reaches
the threshold value specified in Eq.(\ref{Band Gaps}). The fact of
being able to predict band-gaps by means of a micromorphic model is
a novel feature of the introduced relaxed model. Indeed, as it will
be shown in the remainder of this paper, neither the classical micromorphic
continuum model nor the Cosserat and the second gradient ones are
able to predict such band-gaps.

\subsection{The Cosserat model as limit case of the relaxed micromorphic model }

In this subsection we analyze wave propagation in Cosserat-type media
which can be obtained from the proposed relaxed model by letting the
parameter $\mu_{h}$ tend to infinity. Indeed, since the strain energy
density defined in (\ref{KinPot}) must remain finite, when letting
$\mu_{h}\rightarrow\infty$ one must have $\mathrm{Sym}\,\mathbf{P}\rightarrow0$,
hence $\mathbf{P}=\mathrm{skew}\,\mathbf{P}$. Therefore, the strain
energy density becomes the Cosserat energy (see \cite{Neff_JeongMMS08,Jeong_Neff_ZAMM08})
\[
W_{\mathrm{Coss}}\left(\nabla\mathbf{u},\mathrm{skew}\:\mathbf{P}\right)=\mu_{e}\left\Vert \mathrm{sym}\:\nabla\mathbf{u}\right\Vert ^{2}+\frac{\lambda_{e}}{2}\left(\mathrm{tr}\:\nabla\mathbf{u}\right)^{2}+\mu_{c}\left\Vert \mathrm{skew}\left(\nabla\mathbf{u}-\mathbf{P}\right)\right\Vert ^{2}+\frac{\alpha}{2}\left\Vert \mathrm{\mathrm{\, Cur}l}\left(\mathrm{skew}\:\mathbf{P}\right)\right\Vert ^{2}.
\]

\begin{figure}[H]
\begin{centering}
\begin{tabular}{ccccc}
\includegraphics[scale=0.7]{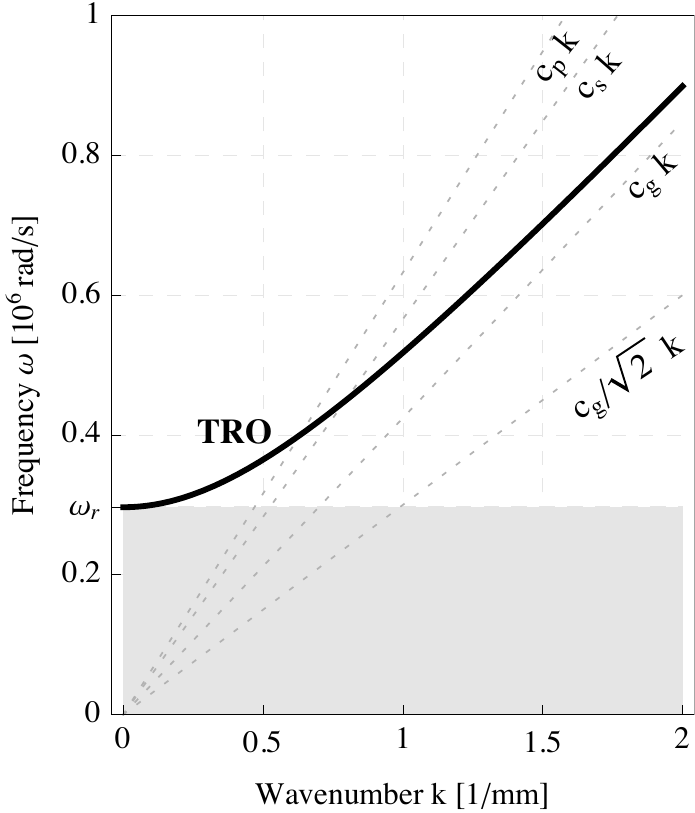}  & \includegraphics[scale=0.7]{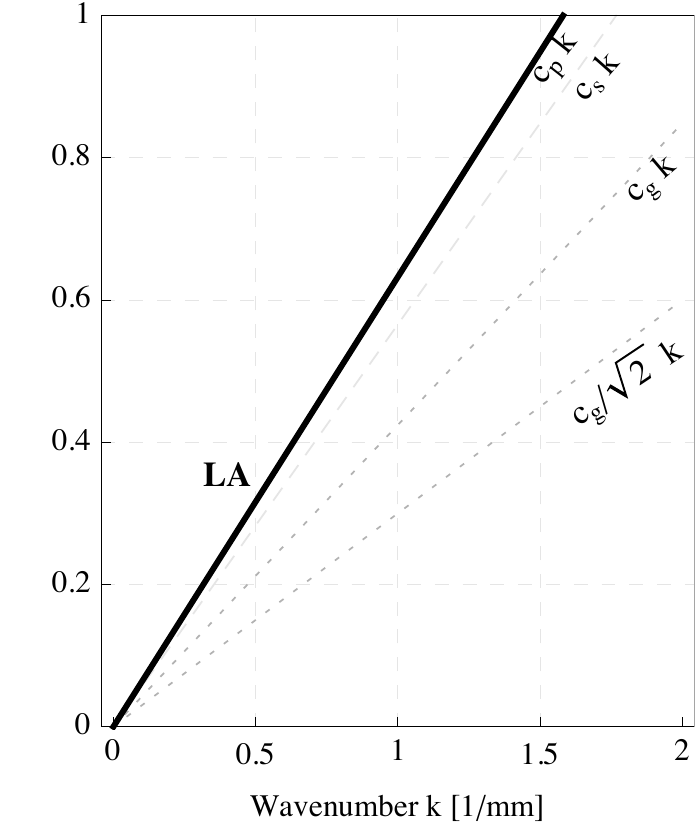} & \includegraphics[scale=0.7]{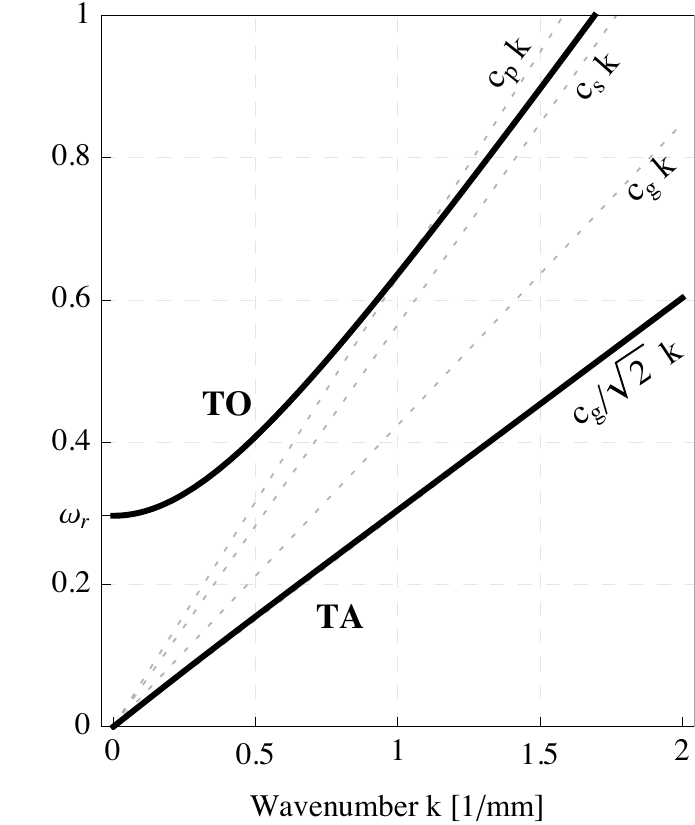} &  & \tabularnewline
(a) & (b) & (c) &  & \tabularnewline
\end{tabular}
\par\end{centering}

\caption{\label{separate-1}Dispersion relations $\omega=\omega(k)$ for the
Cosserat model obtained by letting $\mu_{h}\rightarrow\infty$ in
the relaxed model: uncoupled waves (a), longitudinal waves (b) and
transverse waves (c). TRO: transverse rotational optic, LA: longitudinal
acoustic, TO: transverse optic, TA: transverse acoustic.}
\end{figure}
The dispersion relations obtained for this particular limit case are
depicted in Fig.\ref{separate-1}. It can be immediatly noticed that
all the optic waves with cutoff frequency $\omega_{s}$ do not propagate
anymore since indeed $\omega_{s}\rightarrow\infty$ (see definition
in Eq.(\ref{Definitions})). The same is for the optic longitudinal
wave with cutoff frequency $\omega_{p}$. As for the acoustic waves,
they exist in the limit Cosserat medium, but they do not have horizontal
asymptotes anymore since $\omega_{l}$ and $\omega_{t}$ tend to infinity
as well. We finally end up with a medium in which we can observe one
optic wave ($TRO$) associate to the variable $P_{[23]}$ (micro-rotation)
with cutoff frequency $\omega_{r}$, one acoustic non-dispersive longitudinal
wave ($LA$), one acoustic slightly dispersive transverse wave ($TA$)
and one optic transverse wave ($TO$) with cutoff frequency $\omega_{r}$.
It is easy to remark that no band-gaps can be described in the framework
of the considered Cosserat medium.

The dispersive behavior of Cosserat media shown in Fig.\ref{separate-1}
fits with known results in the literature. Indeed, a direct comparison
with the micropolar medium studied in \cite{EringenBook} (p. 150)
can be made, by simply considering the parameters identification (\ref{Identification1-1}),
(\ref{Identification2-1}). Moreover, Lakes states in \cite{Lakes_Cosserat}
that ``Dilatational waves propagate non-dispersively, i.e. with velocity
independent of frequency, in an unbounded isotropic Cosserat elastic
medium as in the classical case. Shear waves propagate dispersively
in a Cosserat solid (Eringen, 1968). A new kind of wave associated
with the micro-rotation is predicted to occur in Cosserat solids''.
We also remark that the behavior for high frequencies shown in Fig.\ref{separate-1}
coincides with that of acceleration waves in micropolar media, see
\cite{Victor02,Victor2}.

Indeed, the linear Cosserat model is undoubtedly the most studied
generalized continuum model. This does not mean, however, that its
status as a useful model for the description of material behaviour
is unchallenged. Quite to the contrary, it appears that after 40 years
of intensive research, not one material has been conclusively established
as a Cosserat material. We refer to the discussion in \cite{Neff_Cosserat_plasticity05,Neff_Gamm04,Neff_ZAMM05,Jeong_Neff_ZAMM08,Neff_Jeong_bounded_stiffness09,Neff_JeongMMS08,Neff_Jeong_Conformal_ZAMM08}.

By looking at the dispersion behaviour of the linear Cosserat model
and comparing it with the more general micromorphic (and relaxed micromorphic
model) we get a glimpse on why the status of the linear Cosserat model
is really challenging. The equations appear as a formal limit in which
$\mu_{h}\to\infty$, while $\ensuremath{0<\mu_{c}<\infty}$. The process
$\mu_{h}\to\infty$ corresponds conceptually to assume that the substructure
cannot deform elastically, nevertheless the substructures may mutually
interact through the curvature energy, which itself is only involving
the $\mathrm{Curl}$-operator, since the micro-distortion is constrained
to be skew-symmetric.

The usefulness of a geometrically nonlinear Cosserat model, however,
is not in general questioned. Indeed, it has been shown in \cite{Merkel2}
that a Cosserat-type continuum theory can be of use to describe the
experimental behavior of granular phononic crystals. One may avoid
the deficiencies of the linear model and the linear coupling, see
e.g. \cite{Neff_Muench_simple_shear09,Muench07_diss,Neff_Muench_transverse_cosserat08,Neff_Muench_magnetic08,Neff_Biot07}.

\section{The classical micromorphic model: numerical results}

In this section, we show the dispersion relations for a classical
micromorphic continuum.
\begin{figure}[H]
\begin{centering}
\begin{tabular}{ccccc}
\includegraphics[scale=0.7]{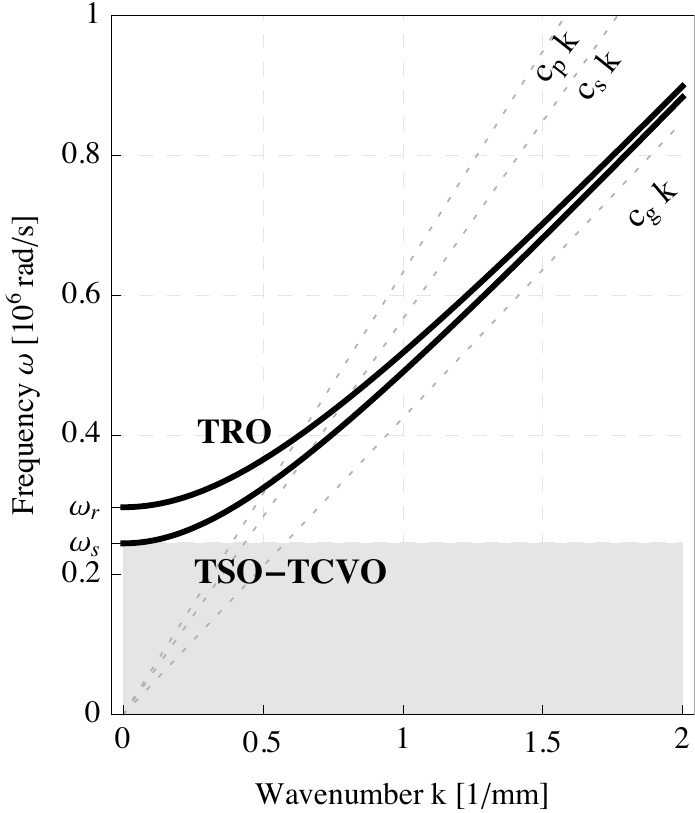}  & \includegraphics[scale=0.7]{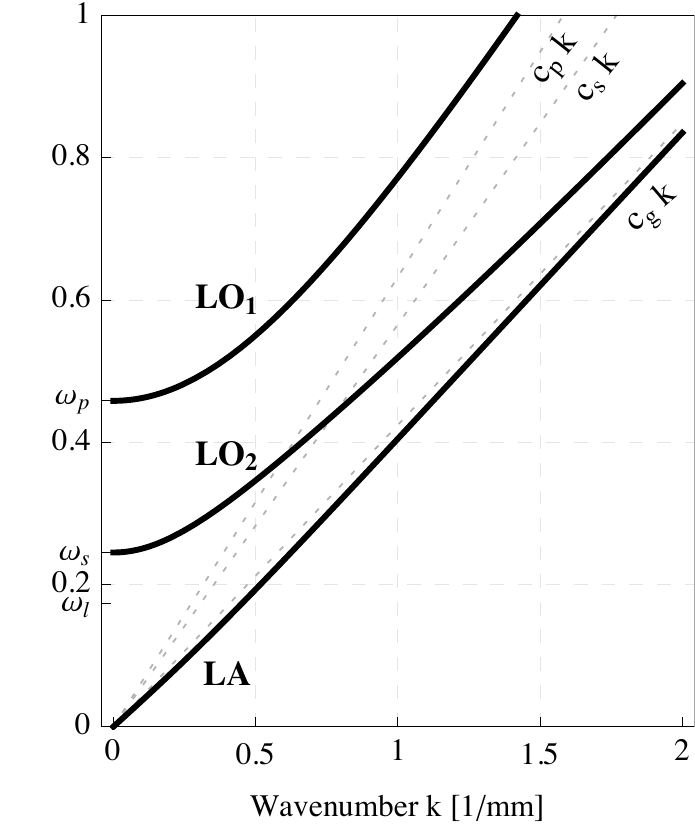} & \includegraphics[scale=0.7]{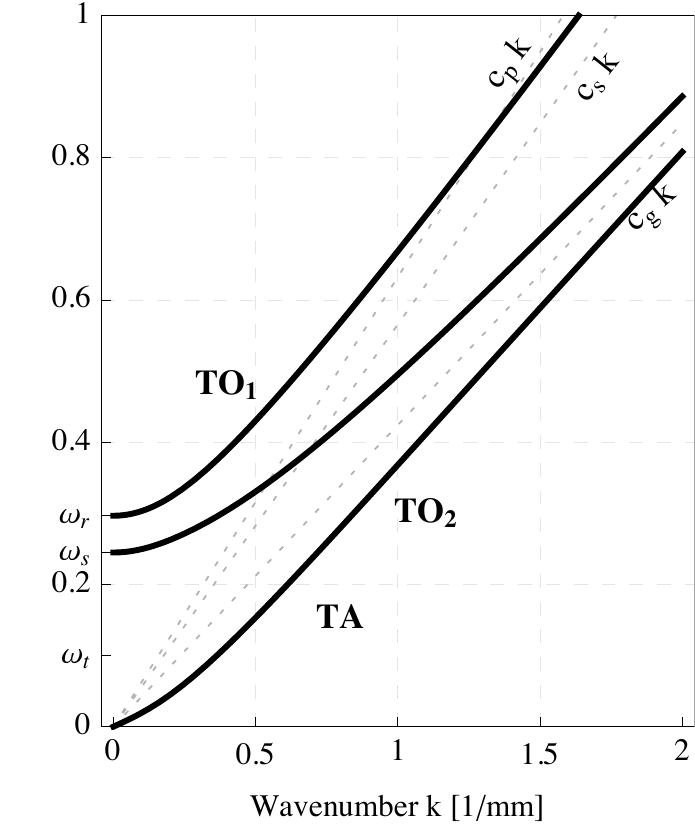} &  & \tabularnewline
(a) & (b) & (c) &  & \tabularnewline
\end{tabular}
\par\end{centering}

\caption{\label{Mindlin}Dispersion relations $\omega=\omega(k)$ for the classical
model: uncoupled waves (a), longitudinal waves (b) and transverse
waves (c) for a classical micromorphic continuum. TRO: transverse
rotational optic, TSO: transverse shear optic, TCVO: transverse constant-volume
optic, LA: longitudinal acoustic, LO$_{1}$-LO$_{2}$: first and second
longitudinal optic, TA: transverse acoustic, TO$_{1}$-TO$_{2}$:
first and second transverse optic.}
\end{figure}
 These dispersion relations are plotted in Fig.\ref{Mindlin}: we
recover a similar behavior with respect to the one qualitatively sketched
in \cite{Mindlin} and \cite{EringenBook}. Indeed, we claim that
the classical micromorphic model associated to the strain energy density
(\ref{KinPot-1-1}) is qualitatively equivalent to the full 18-parameters
Mindlin and Eringen micromorphic model. It is clear from Fig.\ref{Mindlin}
that, even if the behavior of such medium is the same as the one observed
in Fig.\ref{separate} for the relaxed micromorphic medium when considering
small wavenumbers (large wavelengths), the situation is completely
different when considering small wavelengths. Indeed, the first macroscopic
feature that can be observed is that no band-gaps can be forecasted
by a Mindlin-Eringen model due to the fact that no horizontal asymptotes
exist for acoustic waves.

\subsection{A second gradient model as limit case of the classical micromorphic
model }

In this subsection we analyze wave propagation in second gradient
media which can be obtained as a suitable limit case of the classical
micromorphic continua. More particularly, if we let simultaneously
\[
\mu_{e}\rightarrow\infty,\qquad\mu_{c}\rightarrow\infty,
\]
this implies that
\[
\mathrm{sym}\:\mathbf{P}\rightarrow\mathrm{sym}\:\nabla\mathbf{u}\qquad\mathrm{skew}\:\mathbf{P}\rightarrow\mathrm{skew}\:\nabla\mathbf{u}\quad\text{{and\:\ hence}\quad\ensuremath{\mathbf{P}\rightarrow\nabla\mathbf{u}}}.
\]
This means that the energy (\ref{KinPot-1-1}) reduces to

\begin{gather}
W_{2G}\left(\nabla\mathbf{u},\nabla\nabla\mathbf{u}\right)=\mu_{h}\left\Vert \,\mathrm{sym}\:\nabla\mathbf{u}\,\right\Vert ^{2}+\frac{\lambda_{h}}{2}\left(\mathrm{tr}\,\nabla\mathbf{u}\right)^{2}+\frac{\alpha_{g}}{2}\left\Vert \nabla\nabla\mathbf{u}\right\Vert ^{2}\label{Second Gradient}
\end{gather}
which is indeed a second gradient energy for linear-elastic, isotropic
media (see e.g. \cite{2Grad1,2Grad2}). Indeed, governing equations
for second (and higher) gradient continua can be also obtained by
using a simplified kinematics with respect to the one introduced in
this paper and by adopting variational principles (see e.g. \cite{SecGrad1,2Grad2,SecGrad3}).
On the other hand, the fact of obtaining a second gradient model as
the limit case of a micromorphic one can have many advantages either
with respect to numerical efficiency and to physical interpretation
of the boundary conditions (see e.g. \cite{AngelaMicromorphic,Microm1,Microm2}).
\begin{figure}[H]
\begin{centering}
\begin{tabular}{ccccc}
\includegraphics[scale=0.7]{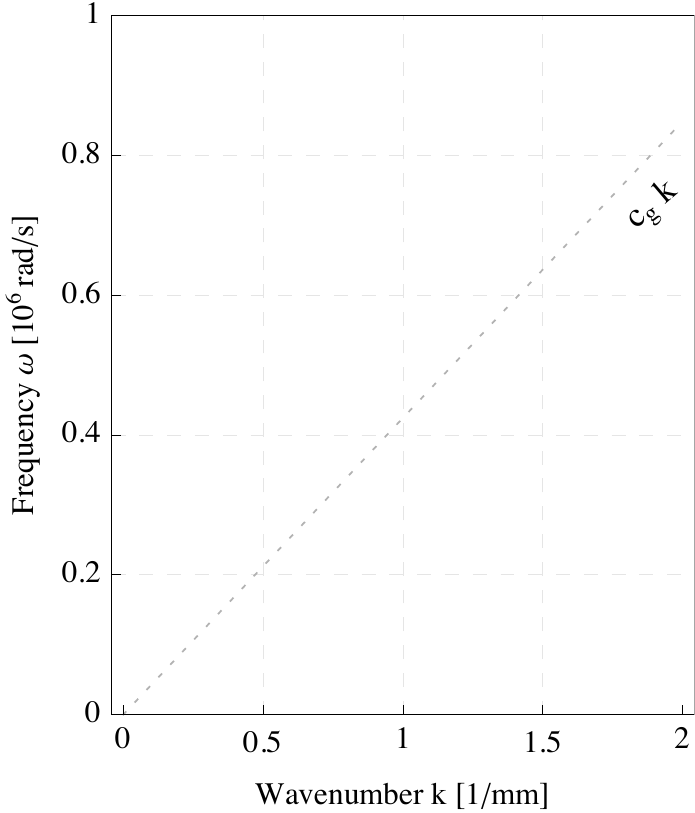}  & \includegraphics[scale=0.7]{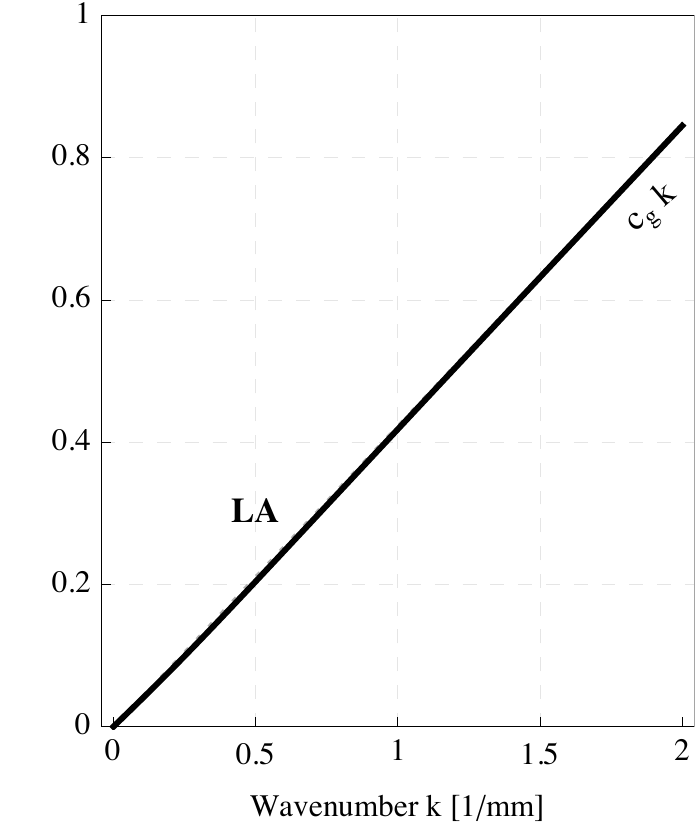} & \includegraphics[scale=0.7]{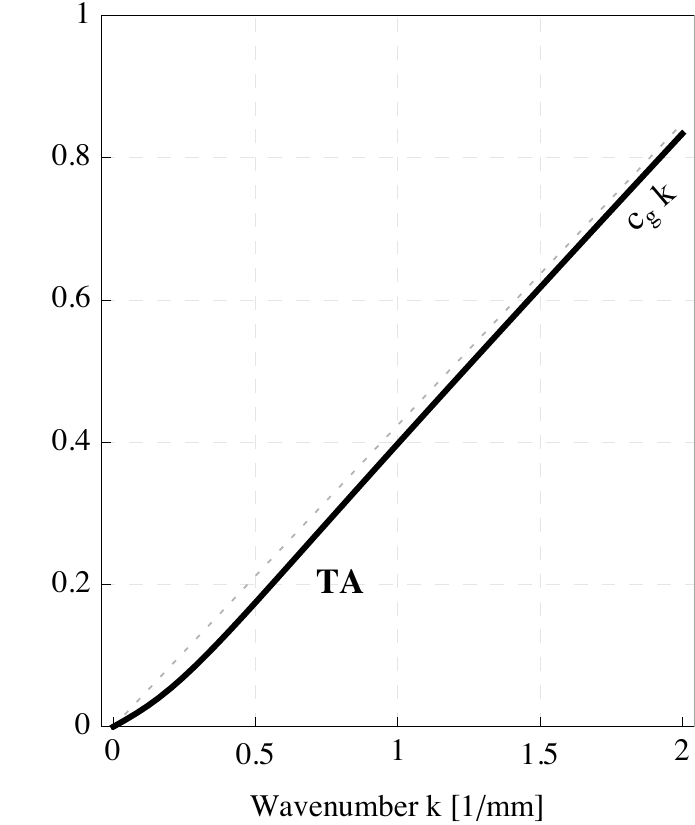} &  & \tabularnewline
(a) & (b) & (c) &  & \tabularnewline
\end{tabular}
\par\end{centering}

\caption{\label{2Grad}Dispersion relations $\omega=\omega(k)$ for the second
gradient model obtained as limit case of the classic model by letting
$\mu_{e}\rightarrow\infty$ and $\mu_{c}\rightarrow\infty$: uncoupled
waves (a), longitudinal waves (b) and transverse waves (c) for a second
gradient continuum. LA: longitudinal acoustic, TA: transverse acoustic.}
\end{figure}
 It is immediate to notice from Fig.\ref{2Grad} that no coupled waves
can exist in this limit case, since the only independent kinematical
variable turns to be the displacement $\mathbf{u}$. As for, the longitudinal
and transverse waves, they can only be of acoustic type ($LA$ and
$TA$). This is in agreement with what is known from the literature
(see e.g. \cite{Waves2Grad1,Waves2Grad2}): second gradient media
are such that only acoustic dispersive waves can propagate inside
the medium. In \cite{Waves2Grad1}, Eq. (13), dispersion relations
for planar waves in isotropic second gradient media are presented
associated to a strain energy density of the type (\ref{Second Gradient})
but with a kinetic energy which is simpler than the one used in this
paper and which can be obtained from (\ref{Kinetic}) by setting $\eta=0$.
It can be checked by means of the simple application of a variational
principle that the equations of motion associated to the strain energy
density (\ref{Second Gradient}) and to the kinetic energy $T=\frac{1}{2}\,\rho\left\Vert \,\mathbf{u}_{t}\,\right\Vert ^{2}+\frac{1}{2}\,\eta\left\Vert \,\left(\nabla\mathbf{u}\right)_{t}\,\right\Vert ^{2},$
can be written as
\begin{equation}
\rho\,\ddot{u}_{i}-\eta\,\ddot{u}_{i,jj}=\mu_{h}\left(u_{i,jj}+u_{j,ij}\right)+\lambda_{h}u_{j,ji}-\alpha_{g}u_{i,jkkj}.\label{Motion}
\end{equation}
It can also be cheked that the dispersion relations for longitudinal
and transverse planar waves (obtained by studying the wave-form solution
of Eq. (\ref{Motion}) when setting $i=1$ and $i=2,3$ respectively)
are respectively given by
\[
\omega=\sqrt{\frac{\alpha_{g}\, k^{4}+(2\,\mu_{h}+\lambda_{h})\, k^{2}}{\rho+\eta\, k^{2}}},\qquad\qquad\omega=\sqrt{\frac{\alpha_{g}\, k^{4}+\mu_{h}\, k^{2}}{\rho+\eta\, k^{2}}},
\]
which indeed correspond to the two acoustic waves depicted in Fig.\ref{2Grad}.
It is immediate to check that for vanishing wavenumber $k$ the frequency
is always vanishing, which means that only acoustic waves can propagate
in such kind of media, optic waves are hence forbidden \textit{a priori}
in this kind of models. Moreover, it is possible to remark that the
frequency goes to infinity when the wavenumber tends to infinity (no
possibility of horizontal asymptotes). Finally, we notice that standing
waves can also be present in second gradient media since purely imaginary
wavenumbers may exist which give rise to real frequencies. Indeed,
when replacing an imaginary wavenumber in the wave-form $e^{i(kX-\omega t)}$
it is immediate to see that the solution is an exponential decaying
in space. We can summarize by saying that, when considering second
gradient media, one has that, for any value of the frequency, propagative
acoustic waves and standing waves exist. It can also be remarked that
the inertial term $\frac{1}{2}\,\eta\left\Vert \,\left(\nabla\mathbf{u}\right)_{t}\,\right\Vert ^{2},$
plays a role on the possibility of changing the concavity of the dispersion
relations. Indeed, if the microscopic density $\eta$ tends to zero,
then it is clear that the concavity of the dispersion curves cannot
change. On the other hand, this change of concavity is possible when
$\eta\neq0$.

We explicitly remark that, in the considered second gradient medium
no band-gaps can be forecasted. Nevertheless, it is known that if
surfaces of discontinuity of the material properties are considered,
then reflected and transmitted energy can be strongly influenced by
the value of the second gradient parameter depending on the considered
jump conditions imposed at the interface itself (see e.g. \cite{Waves2Grad1,Waves2Grad2}).

\section{Conclusions}

In this paper we used the relaxed micromorphic model proposed in \cite{NeffRelaxed ,Ghiba}
to study wave propagation in unbounded continua with microstructure.
The quoted relaxed model only counts 6 elastic parameters against
the 18 parameters appearing in Mindlin-Eringen micromorphic theory
(cf. \cite{Mindlin,EringenBook}). Despite the reduced number of parameters,
we claim that the proposed relaxed model is fully able to account
for the description of the mechanical behavior of micromorphic media.
More precisely, we have shown that only the relaxed micromorphic continuum
model with non-vanishing Cosserat couple modulus $\mu_{c}$ is able
to predict frequency band-gaps corresponding to which no wave propagation
can occur. The main findings of the present study can be summarized
as follows:
\begin{itemize}
\item The relaxed micromorphic model (6 parameters and only the term $\left\Vert \,\mathrm{Curl}\:\mathbf{P}\,\right\Vert ^{2}$
appearing in the strain energy) is fully able to describe the main
features of the mechanical behaviour of micromorphic continua, first
of all for what concerns the possibility of describing the presence
of band gaps.
\item A reduced relaxed model (5 parameters) can be obtained from the previous
one by setting the Cosserat couple modulus $\mu_{c}\rightarrow0$.
This simplified model produces a \textit{symmetric Cauchy force stress
tensor} and is still able to describe the mechanical behavior of a
big class of microstructured continua. Nevertheless, this model excludes
\textit{a priori} the presence of band-gaps. This means that it is
not suitable to fully describe the behavior of sophisticated microstructured
engineering materials such as phononic crystals and lattice structures.
However, practically all known engineering materials do not show band
gaps. For these materials the reduced model is our alternative of
choice in the family of micromorphic models.
\item The Cosserat model is obtained as a degenerate limit case of the proposed
relaxed model (the full relaxed model with 6 parameters) when letting
$\mu_{h}\rightarrow\infty$ . The Cosserat model is not able to describe
band-gaps but it introduces a \textit{non-symmetric Cauchy stress
tensor}.
\item The classical continuum model (6 parameters but the full $\left\Vert \,\nabla\mathbf{P}\,\right\Vert ^{2}$
appearing in the strain energy) is qualitatively equivalent to the
full 18-parameters Mindlin-Eringen micromorphic continuum model. This
is true since the classical model is controlling all the kinematical
fields of the full model i.e. it is uniformly pointwise definite.
The classical continuum model is not able to forecast band-gaps.
\item Second gradient theories can be obtained as a limit case of the classical
micromorphic continuum model by letting $\mu_{e}\rightarrow\infty$,
$\mu_{c}\rightarrow\infty$. These theories are not able to account
for the presence of band-gaps.
\end{itemize}
We can conclude that only the 6-parameters relaxed model proposed
in \cite{NeffRelaxed ,Ghiba} and used in this paper to study wave
propagation in microstructured media is able to describe the presence
of band-gaps. These band-gaps are seen to be ``switched on'' by
a unique constitutive parameter, namely the Cosserat couple modulus
$\mu_{c}$. The proposed relaxed micromorphic model is hence suitable
to be used for the conception and optimization of metamaterials to
be used for vibration control.

\section*{Acknowledgements}

I.D. Ghiba acknowledges support from the Romanian National Authority
for Scien- tific Research (CNCS-UEFISCDI), Project No. PN-II-ID-PCE-2011-3-0521.
I.D. A. Madeo thanks INSA-Lyon for the financial support assigned
to the project BQR 2013-0054 \textquotedblleft{}Matériaux Méso et
Micro-Héterogènes: Optimisation par Modèles de Second Gradient et
Applications en In\'{g}enierie.

\end{document}